\documentclass[iop]{emulateapj}
\usepackage{color}

\def\star{GSC 03546-01452~}
\def\starnsp{GSC 03546-01452}
\def\bd{MARVELS-6b~}
\def\bdnsp{MARVELS-6b}
\def\bds{MARVELS-6b's~}

\shorttitle{\bd}
\shortauthors{De Lee et al.}

\begin{document}


\title{Very Low Mass Stellar and Substellar Companions to Solar-Like Stars From MARVELS V: A Low Eccentricity Brown Dwarf from the Driest Part of the Desert, \bd}


\author{
Nathan De Lee\altaffilmark{1,2,3}, 
Jian Ge\altaffilmark{3}, 
Justin R. Crepp\altaffilmark{4}, 
Jason Eastman\altaffilmark{5,6,7}, 
Massimiliano Esposito\altaffilmark{8,9}, 
Bruno Femen\'{\i}a\altaffilmark{8,9},
Scott W. Fleming\altaffilmark{3,10,11,12},
B. Scott Gaudi\altaffilmark{5}, 
Luan Ghezzi\altaffilmark{13,14}, 
Jonay I. Gonz\'alez Hern\'andez\altaffilmark{8,9}, 
Brian L. Lee\altaffilmark{3,15}, 
Keivan G. Stassun\altaffilmark{1,2}, 
John P. Wisniewski\altaffilmark{16}, 
W. Michael Wood-Vasey\altaffilmark{17},
Eric Agol\altaffilmark{15}, 
Carlos Allende Prieto\altaffilmark{8,9}, 
Rory Barnes\altaffilmark{15},  
Dmitry Bizyaev\altaffilmark{18}, 
Phillip Cargile\altaffilmark{1}, 
Liang Chang\altaffilmark{3}, 
Luiz N. Da Costa\altaffilmark{13,14}, 
G.F. Porto De Mello\altaffilmark{19,14}, 
Leticia D. Ferreira\altaffilmark{19,14}, 
Bruce Gary\altaffilmark{1}, 
Leslie Hebb\altaffilmark{1,15}, 
Jon Holtzman\altaffilmark{20}, 
Jian Liu\altaffilmark{3}, 
Bo Ma\altaffilmark{3}, 
Claude E. Mack III\altaffilmark{1}, 
Suvrath Mahadevan\altaffilmark{10,11}, 
Marcio A.G. Maia\altaffilmark{13,14}, 
Duy Cuong Nguyen\altaffilmark{3,21}, 
Audrey Oravetz\altaffilmark{18}, 
Daniel J. Oravetz\altaffilmark{18}, 
Martin Paegert\altaffilmark{1}, 
Kaike Pan\altaffilmark{18}, 
Joshua Pepper\altaffilmark{1}, 
Elena Malanushenko\altaffilmark{18},
Viktor Malanushenko\altaffilmark{18},
Rafael Rebolo \altaffilmark{8,9,22}, 
Basilio X. Santiago\altaffilmark{23,14}, 
Donald P. Schneider\altaffilmark{10,11}, 
Alaina C. Shelden Bradley\altaffilmark{18}, 
Xiaoke Wan\altaffilmark{3}, 
Ji Wang\altaffilmark{3}, 
Bo Zhao\altaffilmark{3}}

\altaffiltext{1}{Department of Physics and Astronomy, Vanderbilt University, Nashville, TN 37235, USA;nathan.delee@vanderbilt.edu}
\altaffiltext{2}{Department of Physics, Fisk University, Nashville, TN, USA}
\altaffiltext{3}{Department of Astronomy, University of Florida, 211 Bryant Space Science Center, Gainesville, FL, 32611-2055, USA}
\altaffiltext{4}{Department of Physics, University of Notre Dame, 225 Nieuwland Science Hall, Notre Dame, IN 46556, USA}
\altaffiltext{5}{Department of Astronomy, The Ohio State University, 140 West 18th Avenue, Columbus, OH 43210, USA}
\altaffiltext{6}{Las Cumbres Observatory Global Telescope Network, 6740 Cortona Drive, Suite 102, Santa Barbara, CA 93117, USA}
\altaffiltext{7}{Department of Physics Broida Hall, University of California, Santa Barbara, CA 93106, USA}
\altaffiltext{8}{Instituto de Astrof\'{\i}sica de Canarias (IAC), E-38205 La Laguna, Tenerife, Spain}
\altaffiltext{9}{Departamento de Astrof\'{\i}sica, Universidad de La Laguna, 38206 La Laguna, Tenerife, Spain}
\altaffiltext{10}{Department of Astronomy and Astrophysics, The Pennsylvania State University, 525 Davey Laboratory, University Park, PA 16802, USA}
\altaffiltext{11}{Center for Exoplanets and Habitable Worlds, Pennsylvania State University, University Park, PA 16802, USA}
\altaffiltext{12}{Space Telescope Science Institute, 3700 San Martin Drive, Baltimore, MD 21218}
\altaffiltext{13}{Observat\'orio Nacional, Rua Gal. Jos\'e Cristino 77, Rio de Janeiro, RJ 20921-400, Brazil}
\altaffiltext{14}{Laborat\'orio Interinstitucional de e-Astronomia - LIneA, Rua Gal. Jos\'e Cristino 77, Rio de Janeiro, RJ 20921- 400, Brazil}
\altaffiltext{15}{Astronomy Department, University of Washington, Box 351580, Seattle, WA 98195, USA}
\altaffiltext{16}{H L Dodge Department of Physics and Astronomy, University of Oklahoma, 440 W Brooks St Norman, OK 73019, USA}
\altaffiltext{17}{Pittsburgh Particle physics, Astrophysics, and Cosmology Center (PITT PACC), Department of Physics and Astronomy, University of Pittsburgh, Pittsburgh, PA 15260, USA}
\altaffiltext{18}{Apache Point Observatory, P.O. Box 59, Sunspot, NM 88349-0059, USA}
\altaffiltext{19}{Universidade Federal do Rio de Janeiro, Observatório do Valongo, Ladeira do Pedro Antonio 43, 20080-090 Rio de Janeiro, Brazil}
\altaffiltext{20}{Department of Astronomy, New Mexico State University, Box 30001, Las Cruces, NM 880033, USA}
\altaffiltext{21}{Department of Physics and Astronomy, University of Rochester, Rochester, NY 14627-0171, USA}
\altaffiltext{22}{Consejo Superior de Investigaciones Cient\'{\i}ficas, Spain}
\altaffiltext{23}{Instituto de Fisica, UFRGS, Porto Alegre, RS 91501-970, Brazil}

\submitted{Accepted by The Astronomical Journal on April 8th, 2013}


\begin{abstract}
We describe the discovery of a likely brown dwarf (BD) companion with
a minimum mass of 31.7 $\pm$ 2.0 $M_{Jup}$ to \star from the MARVELS
radial velocity survey, which we designate as \bdnsp. For reasonable
priors, our analysis gives a probability of 72\% that \bd has a mass
below the hydrogen-burning limit of 0.072 $M_{\sun}$, and thus it is a
high-confidence BD companion.  It has a moderately long orbital period
of $47.8929^{+0.0063}_{-0.0062}$ days with a low eccentricty of
$0.1442^{+0.0078}_{-0.0073}$, and a semi-amplitude of
$1644^{+12}_{-13}$ $\rm m~s^{-1}$.  Moderate resolution spectroscopy
of the host star has determined the following parameters: $T_{\rm eff}
= 5598 \pm 63$, $\log{g} = 4.44 \pm 0.17$, and [Fe/H] = $+0.40 \pm
0.09$. Based upon these measurements, \star has a probable mass and
radius of $M_*= 1.11 \pm 0.11~M_\sun$ and $R_*= 1.06 \pm 0.23~R_\sun$
with an age consistent with less than $\sim$6 Gyr at a distance of 219
$\pm$ 21 pc from the Sun. Although \bd is not observed to transit, we
cannot definitively rule out a transiting configuration based on our
observations. There is a visual companion detected with Lucky Imaging
at 7.7\arcsec~from the host star, but our analysis shows that it is not
bound to this system. The minimum mass of \bd exists at the minimum of
the mass functions for both stars and planets, making this a rare
object even compared to other BDs. It also exists in an underdense
region in both period/eccentricity and metallicity/eccentricity space.
\end{abstract}





\section{Introduction}
Radial velocity (RV) surveys have provided a wealth of exoplanet
discoveries around sun-like stars in recent years (California Planet
Survey, \citet{how10}, Lick-Carnegie Exoplanet Survey, \citet{hag10}; CORALIE
survey, \citet{udr00}; and the HARPS survey, \citet{may03} to name a
few), but they have not found a correspondingly large number of brown
dwarf (BD) companions \citep{rei08}. BD companions lie on the mass
spectrum between planets and stars and are defined as being between
13$M_{Jup}$ and 75.5 $M_{Jup}$ (based on the deuterium and hydrogen
fusion limits) \citep{cha00,spi11}. The lack of BDs within 3 AU of their
host star was first recognized in \citet{mar00}, and is known as the
BD ``desert''. This result is unlikely to be due to observational bias
because the RV semiamplitudes of BD are many hundreds to a few
thousand meters per second, which is easily detectable by these
surveys \citep{pat07}. Thus the lack of BD at close to moderate
distances from their respective host stars points to an explanation
based in BD formation mechanisms.

Although the formation of low-mass companions (planets through low
mass stars) to sun-like stars is still an area of active research, in
overview there are two main mechanisms: planets form from a
protoplanetary disks and stellar companions [in a similar range of
  separations] from molecular cloud fragmentation. Given the mass
range of BD companions they could form through either mechanism (or
both). Understanding the origin of the BD desert can put major
constraints on the upper mass limit for companion formation in
protoplanetary disks, and a lower mass limit on formation via
fragmentation.

Recent efforts to quantify the frequency of companions as a function
of mass in the BD desert have found that the overall frequency of BD
companions at close to moderate distances from their host star ($\le$
10 AU) is less than $<1\%$ \citep{gre06}, and more recently $0.6\%$
\citep{sah11}. This value is low compared to $\sim7\%$ for planetary
companions \citep{udr07} and $\sim13\%$ for stellar companions in a
similar range of separations \citep{duq91,hal03}. \citet{gre06} went a
step further and defined the driest part of the desert to be where
there was a minimum in the number of companions per unit interval in
log mass; they found this position to be at a companion mass of
$31^{+25}_{-18}M_{Jup}$.

This is the fifth paper in this series looking at low-mass companions
to sun-like stars from the third generation of the Sloan Digital Sky
Survey (SDSS-III; \citet{eis11}) Multi-Object APO Radial Velocity
Exoplanet Large-Area Survey \citep[MARVELS;][]{ge08,ge09a,ge09b}. The
primary goal of this series is to provide a detailed set of
well-characterized companions with minimum masses and separations in
or near the BD desert, which can be used by future
meta-analyzes. Ultimately, this is the sort of groundwork that must be
done in order to understand the extent and aridity of the BD desert.

The MARVELS survey measured radial velocities of 3,300 unique FGK type
stars. MARVELS is a large survey looking for RV companions around
bright stars ($7.6 \leq V \leq 12$) with periods below 2 years with
well characterized biases; see \citet{lee11} for a description of the
survey design. Other papers in this series \citep[Mack et al. 2013,
  in press; Jiang et al. 2013, submitted]{fle10,fle12,wis12,ma13a}
have helped fill in our understanding of the BD desert, and provided
warnings to some of the pitfalls inherent in these analyzes.

We will discuss observations of the star \starnsp, which has a companion
with a period of $\sim47$ days and with a minimum mass of $31.7 \pm 2.0 M_{Jup}$ placing
it near the most ``arid'' region of the BD desert. In Section 2 we discuss
the photometric and spectroscopic observations and basic data
processing. In Section 3 we discuss the analysis of this data. Section
4 contains a discussion of the results and places \bd within the
larger context. Section 5 summarizes our results.

\begin{figure}[tbp]
\begin{flushright}
\includegraphics[width=\linewidth]{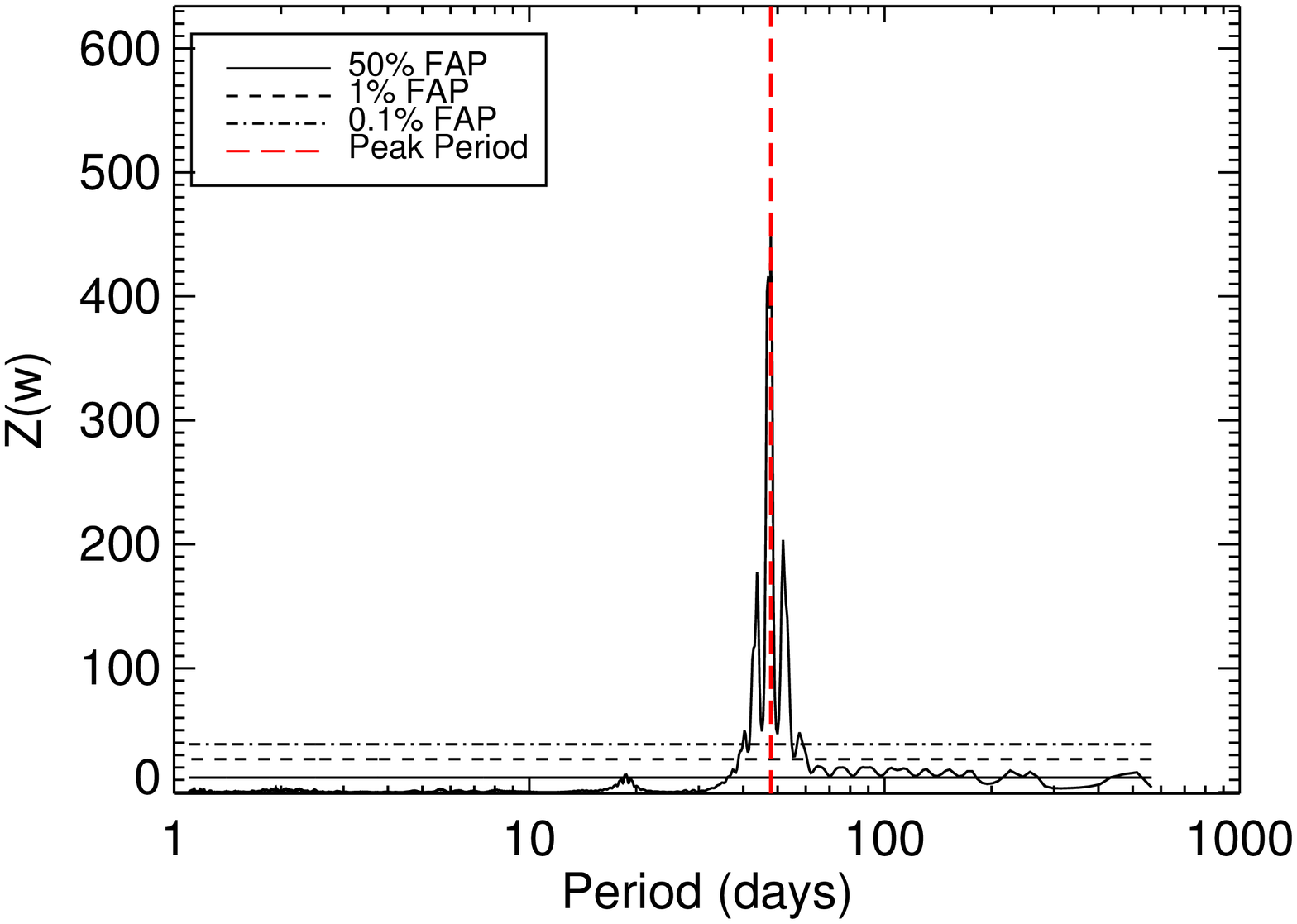}
\includegraphics[width=.98\linewidth]{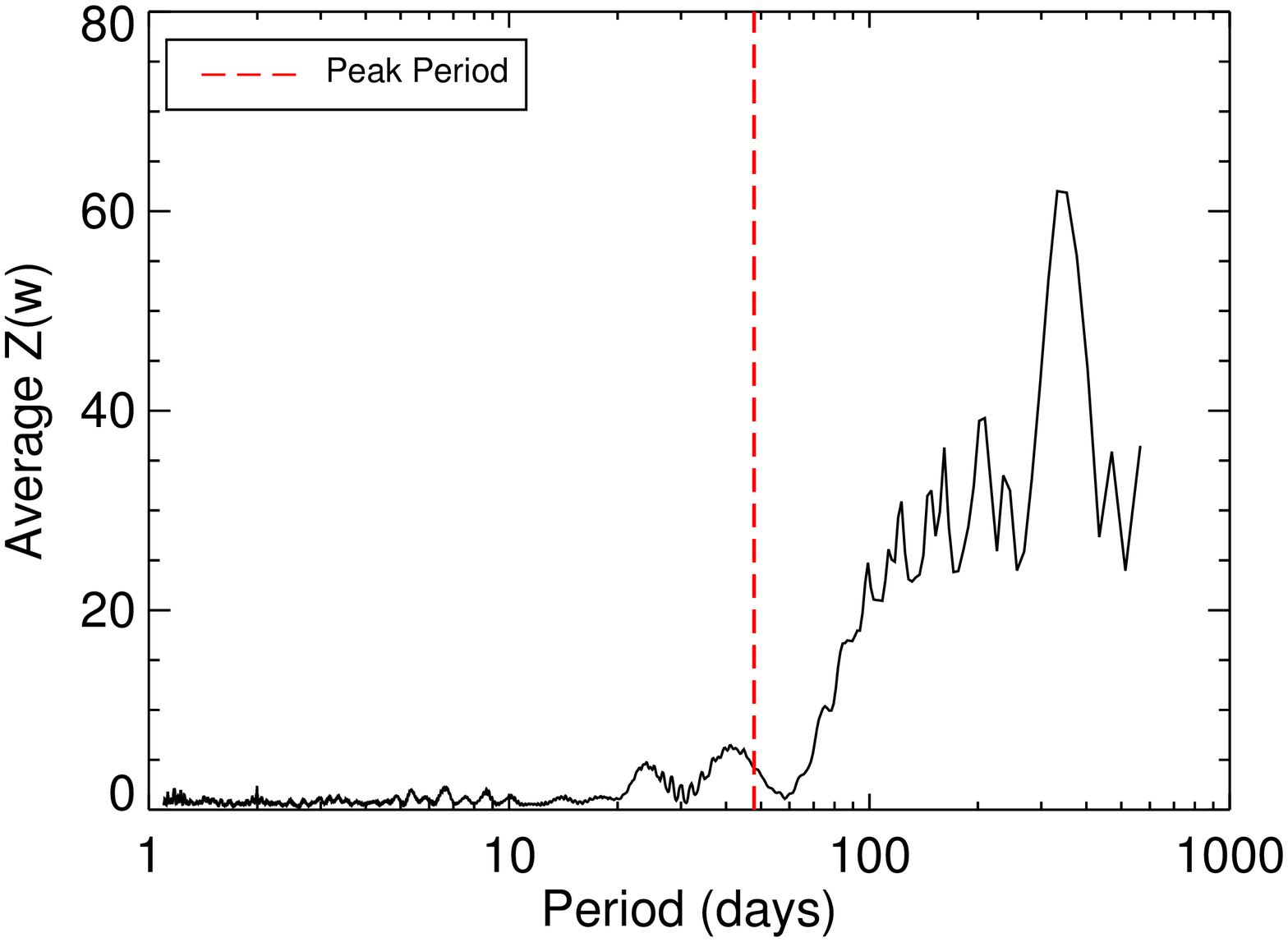}
\end{flushright}
\caption{Top: The original detection Lomb-Scargle periodogram for \star
  based on the MARVELS data. The two RV points with the lowest
  spectrum flux were removed for this initial periodogram (as marked
  in \mbox{Table \ref{rvtab}}). The peak in the periodogram (marked by
  the red dashed line) is quite significant and is at 47.842 days,
  close to the final adopted value of 47.893 days for the combined
  MARVELS and TNG/SARG data. Bottom: A combined periodogram for all 60
  stars on the MARVELS plate. The highest power point in each
  frequency bin was removed, and the remaining were averaged. The red
  dashed line shows the peak period from the periodogram on the
  left.There is no obvious systematic peak on the plate periodogram at
  the peak period of the \star periodogram.\label{detectfig}}
\end{figure}

\section{Observations and Data reduction}
The initial detection and orbitial characterization of \bd came from
the MARVELS Survey RV data. Once it became a strong candidate further
RV, photometric, time-series photometric, and spectroscopic data were
taken to confirm this candidate. Each of these data sets will be
discussed below.
\subsection{Initial Identification of MARVELS Candidates}
\subsubsection{Survey Summary}
\label{msurvey}
MARVELS is a multi-epoch radial velocity survey designed to
detect radial velocity companions around FGK type stars in a magnitude
range of ($7.6 \leq V \leq 12$). It uses a dispersed fixed-delay
interferometer \citep[DFDI;][]{ge09a} on the SDSS 2.5m telescope
\citep{gun06}.  The DFDI method was introduced for use in a
multi-object RV survey by \citet{ge02a}. A single object version of a
DFDI instrument was successfully used to detect a hot Jupiter around
HD 102195 \citep{ge06}.  The DFDI instrument principle was described
by \citep{ge02a,ge02b,ers03,van10,wan11}.  The MARVELS interferometer
delay calibrations were described in \citet{wan12a,wan12b}.

Each MARVELS observation consists of 60 stars spread
across 120 spectra (2 spectra for each star). The MARVELS survey
started in Oct 2008 and ran through July 2012. It was divided up into
two observing sets: year 1-2 fields which were observed from Oct 2008
till Jan 2011 and year 3-4 fields which were observed from Jan 2011
till July 2012. 

The year 1-2 data set consisted of 43 unique fields (11 of which are
in the \emph{Kepler} [\citealp{bor10}]) field of view. Year 3-4 field set
contained 13 fields (with 1 field overlapping the year 1-2 field set.)
This leads to a total of 3,300 unique stars with $>18$ epochs.
\begin{deluxetable}{rrrc}
  \tabletypesize{\footnotesize}
  \tablewidth{0pt}
  \tablecaption{Summary of Radial Velocity Data\label{rvtab}}
  \tablehead{\colhead{BJD} 
    & \colhead{RV} 
    & \colhead{RV err} 
    & \colhead{Source}\\
    \colhead{(days)}
    & \colhead{($\rm m~s^{-1}$)}
    & \colhead{($\rm m~s^{-1}$)}
    & \colhead{\nodata}  
}
\startdata
2454956.915966 & 2039.20 & 72.10 & MARVELS\tablenotemark{a}\\ 
2454957.946510 & 2150.89 & 43.11 & MARVELS\\
2454958.912457 & 2206.71 & 39.27 & MARVELS\\
2454959.927110 & 2291.25 & 36.06 & MARVELS\\
2454962.924107 & 2568.84 & 36.54 & MARVELS\\
2454963.890111 & 2559.01 & 48.00 & MARVELS\\
2454964.918085 & 2529.69 & 39.97 & MARVELS\\
2454965.910038 & 2579.81 & 43.36 & MARVELS\\
2454984.835892 & -797.93 & 41.72 & MARVELS\\
2454986.892198 & -736.07 & 43.95 & MARVELS\\
2454988.905306 & -444.95 & 40.44 & MARVELS\\
2454990.799823 & -154.83 & 64.34 & MARVELS\\
2454993.804470 &  255.99 & 43.11 & MARVELS\\
2454994.793741 &  543.09 & 49.32 & MARVELS\\
2454995.898027 &  594.44 & 43.11 & MARVELS\\
2455014.802962 & 2438.46 & 51.41 & MARVELS\\
2455020.845356 & 1639.58 & 39.45 & MARVELS\\
2455021.851131 & 1381.28 & 50.68 & MARVELS\\
2455023.835931 &  792.32 & 53.52 & MARVELS\\
2455024.788660 &  463.02 & 51.78 & MARVELS\\
2455484.614289 & 2165.40 & 59.70 & MARVELS\\
2455436.578251 &  1901.69 & 24.22 & TNG/SARG\\
2455460.467627 & -841.33 & 4.49 & TNG/SARG\\
2455460.489293 & -833.73 & 4.44 & TNG/SARG\\
2455460.511619 & -836.81 & 4.76 & TNG/SARG\\
2455521.551045 &  722.14 & 88.25 & MARVELS\tablenotemark{b}\\
2455725.687910 &  1962.38 & 6.86 & TNG/SARG\\
2455760.505697 &   66.39 & 9.99 & TNG/SARG\\
2455760.694598 &  98.71 & 5.87 & TNG/SARG\\
2455791.532669 &  122.53 & 5.96 & TNG/SARG\\
2455791.583907 &  82.56 & 5.52 & TNG/SARG\\
2455844.434368 & -834.69 & 5.08 & TNG/SARG\\
\enddata
\tablenotetext{a}{Second lowest flux spectrum}
\tablenotetext{b}{The lowest flux spectrum}
\end{deluxetable}



\subsubsection{Radial Velocity Analysis software}
The large number of RV targets required development a software package
to easily display and characterize the radial velocity curves from the
MARVELS survey. One of the primary tools used in this process is an
IDL-based Keplerian model fitter known as MPRVFIT\footnote{http://www.vanderbilt.edu/AnS/physics/vida/mprvfit.htm}. MPRVFIT starts with
the input of RV data in the form of Julian date, RV,
and RV error (the Julian date and RVs can be of any type that is
appropriate to the analysis). In the case of MARVELS, the data from
the two beams have been combined as described in \citet{fle10}. The
software then applies a modified version of the Lomb-Scargle
periodogram \citep{lom76,sca82} described in \citet{cum04}. False
alarm probabilities were assigned to the highest peaks using the
formulation of \citet{bal08}. Once a number of high probability peaks
are identified, the frequency space between the peak and the next
nearest frequency point in the periodogram is sub-divided into 10
frequency steps (on both sides of the peak). A Keplerian model is then
fit to those frequencies using MPFIT \citep{mar09}. MPFIT is a
Levenberg-Marquardt non-linear least squares fitter implemented in
IDL. The $\chi^2$ statistic is determined for each fit, and the
best chi-squared fit of the grid is retained. Finally, the best fit
models are plotted with the data points for easy reference. For the
MARVELS survey, the two best fits for each target were displayed, and
the candidates were chosen based on the folded and unfolded radial
velocity curves, as well as the significance of the periodogram. An
example of this periodogram for \star can be found in Figure
\ref{detectfig}.

\subsection{Radial Velocities}
\subsubsection{SDSS-II MARVELS Radial Velocities}
Differential RV observations of \star were
acquired during the first two years of the SDSS-III MARVELS survey. A
total of twenty-two observations were obtained over the course of 565
days. As discussed in Section \ref{msurvey}, the MARVELS survey uses
the DFDI technique which introduces an interferometer into the light
path. As a result, each 50 minute observation includes two fringed
spectra, or beams, one from each arm of the interferometer. Each
spectrum spans a wavelength range roughly 500-570 nm with a
R$\sim$12,000 resolution. Each beam is processed individually through
the MARVELS pipeline following the methods described in \citet{lee11}.

The formal errors derived from the MARVELS pipeline are known to be
underestimates of the true error. This systematic underestimate can be
partially corrected for by using the fact that 60 stars (120 beams)
are taken in each observation. Following the method outlined in
\citet{fle10}, it is expected that most stars in an observation plate
are radial velocity stable, so the median RMS is a reasonable estimate
of the systematic errors. A quality factor (QF) is derived for each
beam, which is the ratio of the radial velocity RMS to the median
formal error bar, and the median QF is found for the plate. The formal
errors are multiplied by this QF (2.334 for \starnsp) resulting in the
error bars used for the analysis of these observations. During the RV
analysis, discussed in Section \ref{rvanalysis}, it was determined
that these scaled error bars were themselves overestimated, leading to
a reduction of factor of 0.5794. The final differential RV
measurements and final scaled error bars (approximately 40\% larger
than the formal uncertainties) for \star are presented in
Table \ref{rvtab}.

\subsubsection{TNG Differential Radial Velocities}
Once \star was determined to be a candidate for additional investigation,
spectroscopic observations were conducted with the Spettrografo Alta
Risoluzione Galileo (SARG) \citep{gra2001} on the 3.6m Telescopio
Nazionale Galileo (TNG) located at Roque de Los Muchachos Observatory
(ORM). All the spectra were acquired using the same instrumental
configuration: a slit with a sky-projected width of $0.8{\arcsec}$ achieving a
resolving power of $R = 57000$; a yellow cross-dispersing
grism providing the wavelength range $462 < \lambda < 792$ nm.

A total of 15 spectra were taken (11 with and 4 without the iodine
cell inserted in the light path). One spectrum with the iodine cell
proved to have too low S/N, leaving only 10 spectra suitable for
differential velocity measurements. The spectra were processed using
the standard IRAF\footnote{IRAF is distributed by the National Optical
  Astronomy Observatories, which are operated by the Association of
  Universities for Research in Astronomy, Inc., under cooperative
  agreement with the National Science Foundation.} Echelle reduction
packages. The S/N per resolution element at 550 nm ranges from 50 to
130. The spectra without I2 lines served as a reference for measuring
the relative radial velocities of the 10 spectra with superimposed I2
absorption lines. The technique adopted to derive RV values is
described in \citet{mar1992}. For details on how the technique was
implemented on SARG spectra, see \citet{fle12}. As is the case with
the MARVELS RV data, the formal uncertainties for the TNG/SARG data
were overestimated, and so the error bars were thus reduced by a
factor of 0.2224. The final differential RV and reduced error bars are
listed in Table \ref{rvtab}.
\subsubsection{TNG Absolute Radial Velocities}
All 15 spectra taken with the TNG/SARG instrument were used to
calculate absolute RVs. The stellar spectra were cross-correlated with
a high resolution solar
spectrum\footnote{http://bass2000.obspm.fr/solar\_spect.php}. The
cross-correlation was done using the SARG red CCD spectrum with a
wavelength coverage of 6200-8000 \AA.  Due to the non-simultaneity of
the ThAr wavelength calibration exposures it is possible the slit
illumination varied between the science spectra and the calibration
spectra, which could result in the RV measurement being affected by
systematic errors. To partially mitigate this error source we
calculated the cross-correlation function of the telluric lines
\citep{gri73} around 6900 \AA~with a numerical mask. This exercise
results in a RV correction of a few hundred $\rm m~s^{-1}$. This
correction plus the barycentric correction were applied to the radial
velocities and the resulting values are shown in Table \ref{absrvtab}.

\begin{deluxetable}{rrr}
  \tabletypesize{\small}
  \tablewidth{0pt}
  \tablecaption{SARG/TNG Absolute Radial Velocities\label{absrvtab}}
  \tablehead{\colhead{BJD} 
    & \colhead{RV}
    & \colhead{RV Err} \\
    \colhead{days}
    & \colhead{($\rm km~s^{-1}$)}
    & \colhead{($\rm km~s^{-1}$)}
} 
\startdata
2455436.550647 & -12.63&0.13\\
2455436.578251 & -12.73&0.13\\
2455460.467627 & -15.43&0.13\\
2455460.489293 & -15.35&0.13\\
2455460.511619 & -15.41&0.13\\
2455698.630462 & -15.28&0.13\\
2455698.654826 & -15.30&0.13\\
2455725.687910 & -12.48&0.13\\
2455760.505697 & -14.39&0.13\\
2455760.694598 & -14.32&0.13\\
2455791.532669 & -14.36&0.13\\
2455791.583907 & -14.22&0.13\\
2455791.610041 & -14.43&0.13\\
2455791.635364 & -14.52&0.13\\
2455844.434368 & -15.48&0.13\\
\enddata
\end{deluxetable}

To validate this method, observations of the RV standard star HD3765
were used. The standard was observed at 8 epochs with a mean result of
-63,155 $\rm m~s^{-1}$ $\pm$ 128 $\rm m~s^{-1}$. This result is within a
1-$\sigma$ agreement with the mean 34 observations taken from the
ELODIE archive of -63,286 $\pm$ 49 $\rm m~s^{-1}$. For these
observations, the systematics caused by the non-simultaneity of the
ThAr wavelength calibration is the dominant error source, so we adopt
$\pm$ 128 $\rm m~s^{-1}$ for the error in these measurements.

\subsection{High-Resolution Spectrum for Stellar Classification}
\label{apospec}
Two R $\sim$31,500 optical ($\sim$3600–-10,000 \AA) spectra of \star
were obtained on UT 2010 June 20 with the Apache Point Observatory 3.5
m telescope and ARC Echelle Spectrograph \citep[ARCES;][]{wan03} to
enable accurate characterization of stellar fundamental
parameters ($T_{\rm eff}$, $\log{g}$, and [Fe/H]). The two spectra were obtained using the default $1.6$ x
$3.2$ arcsec slit and an exposure time of 1200 s for each spectra (for
a combined exposure time of 2400 s). A ThAr lamp exposure was obtained
after each of these integrations to facilitate accurate wavelength
calibration. The data were processed using standard IRAF
techniques. Following heliocentric velocity corrections, each order
was continuum-normalized, and the resultant continuum normalized data
from each observation were averaged. The final spectrum yielded a
signal-to-noise ratio (S/N) of approximately 110 per resolution
element in the region around 6000 \AA.
\subsection{Photometry}
\subsubsection{HAO Absolute Photometry}
We used the Hereford Arizona Observatory (HAO), a private facility in
southern Arizona (observatory code G95 in the IAU Minor Planet Center),
to measure multi-band, absolute photometry of \starnsp.  HAO employs a
14-inch Meade Schmidt-Cassegrain (model LX200GPS) telescope,
fork-mounted on an equatorial wedge located in a dome. The
telescope's CCD is an SBIG ST-10XME with a KAF-3200ME detector. A
10-position filter wheel accommodates SDSS and Johnson/Cousins filter
sets. \star and standard stars were observed with Johnson B and SDSS u', g', r', i'
filters \citep{fuk96}. For the Johnson-Kron-Cousins bands, standard stars are taken
from the list published by \citet{lan07} and \citet{lan09}. For the
SDSS bands, standard stars are taken from the list published by
\citet{smi02}. Observations were conducted on three dates:
24 and 25 April, 2010 and on 7 May, 2010. Between 23 and 75
standard stars (Landolt and SDSS) were used to establish the
transformations to the standard photometric systems.

\subsubsection{Allegheny Lightcurve}
We obtained 44 nights of observations of \star with the 16-inch
Keeler RCX-400 Meade telescope at the Allegheny Observatory,
University of Pittsburgh.  The observations span 18 months from May
2010 through November 2011 and were all made through a Cousins R
filter onto an SBIG KAF-6303E/LE 2048x3072 pixel CCD with a pixel
scale of 0.57\arcsec~$\rm pixel^{-1}$ for a total FoV of 19.5\arcmin~by
29.2\arcmin.  Exposure times ranged from 30--150 seconds depending on
conditions, with a median exposure time of 75 seconds.

Standard bias and dark subtraction and flatfield calibrations were
preformed on the raw data, and the images were astrometrically calibrated
to the 2MASS Point-Source Catalog \citep{skr06}. Typical seeing
conditions were between 3\arcsec--4\arcsec~with a median seeing of
3.5\arcsec. Photometry was accomplished using a 10-pixel (5.7\arcsec) circular
aperture and subtractring the estimated sky flux based on a 15--20
pixel (8.6\arcsec--11.4\arcsec) radius sky annulus.  Typical
1-$\sigma$ uncertainties were 5 millimag. Relative photometry was calculated
relative to the average counts from two reference stars in the image
with J2000 coordinates: (1): 19:11:34.733 +48:34:54.77; and (2) 19:11:56.335
+48:22:55.38.

\subsubsection{SuperWASP Lightcurve}
The SuperWASP survey \citep{pol06} is a wide-angle transiting planet
survey that monitors the brightness of millions of stars. The survey uses a visual broad
band filter that covers the wavelength range from 400-700nm. For details on reduction
techniques and survey design please consult \citet{pol06}. The
SuperWASP photometry for \star consists of 8823 points spanning just
over four years from May 2004 to Aug 2008.

\subsection{High Spatial Resolution Imaging}
Two high spatial resolution imaging campaigns of \star were undertaken
in order to help rule out false positives and to look for hierarchical
structure. In particular, these observations were conducted to search
for visual companions at large separations that could influence our
spectroscopic results, or that might be bound tertiary companions to
\starnsp. The adaptive optics imaging and lucky imaging runs are
complimentary with the adaptive optics focusing near the star, and
lucky imaging covering out to larger distances from the star.

\begin{figure}[tbp]
\plotone{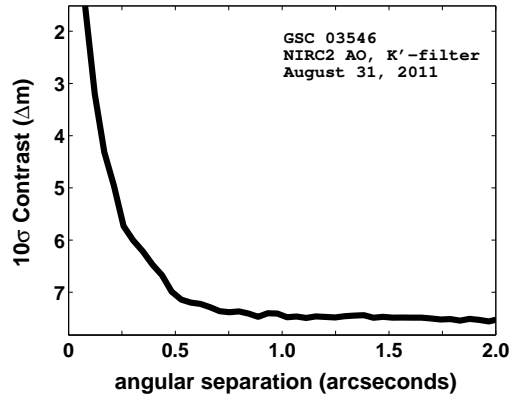}
\caption{Contrast generated by NIRC2 AO observations of \star. Off-axis
  sources with relative brightness $\Delta \rm K'=7$ are ruled out at
  10-$\sigma$ for angular separations beyond $\approx0.5
  \arcsec$.\label{contrast}}
\end{figure}

\subsubsection{Adaptive Optics Imaging}
We also acquired adaptive optics (AO) observations of \star to
assess its multiplicity at wide separations. Images were obtained on
31 Aug., 2011 UT using NIRC2 (PI: Keith Matthews) at the 10m Keck II
telescope \citep{wiz00}.  \star (V=11.7) is sufficiently bright
to serve as its own natural guide star.  Our observations consist of 9
dithered images (10 coadds per frame, 0.5s per coadd) taken with the
$K'$ filter ($\lambda_c=2.12$ $\mu$m). We used NIRC2's narrow camera
setting, which has a plate scale of 10 mas $\rm pixel^{-1}$, to provide fine
spatial sampling of the instrument point-spread function. Raw frames
were processed by cleaning hot pixels, flat-fielding, subtracting
background noise from the sky and instrument optics, and aligning and
coadding the results. No off-axis sources were noticed in individual
frames or the final processed image. Figure \ref{contrast} shows the
contrast levels generated by the observations. Our diffraction-limited
images rule out the presence of companions down to $\Delta K'=5.7,
7.1, 7.5$ mags at angular separations of 0.25", 0.5", and 1.0"
respectively.

\begin{figure}[tbp]
\includegraphics[width=\linewidth]{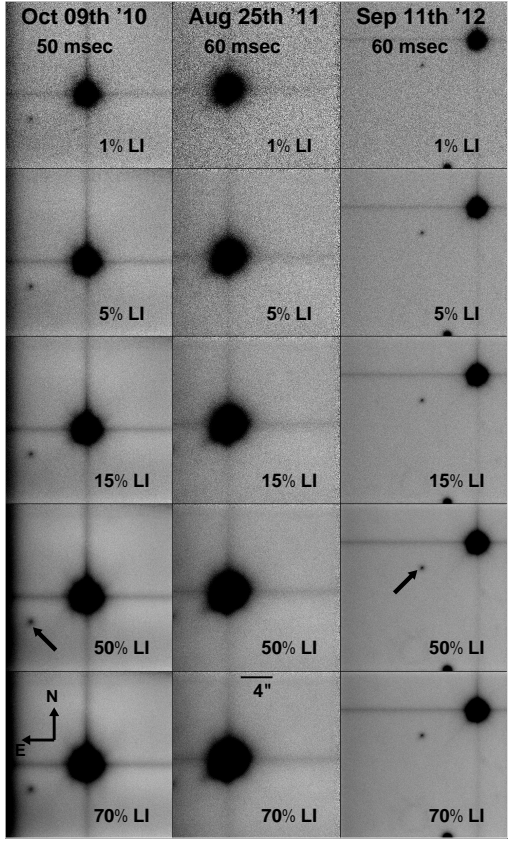}
\caption{\star was observed on 9 Oct., 2010, 25 Aug., 2011, and 11
  Sep., 2012 with LI from the Fastcam imager. The mosaic shows the
  final LI combined images from each run using different
  thresholds. LI detected a previously known visual companion
  7.7\arcsec~away from \star (marked with an arrow) in the 9 Oct.,
  2010 and 11 Sep., 2012 runs. The companion was not detected on the 25 Aug.,
  2011 night because it was not in the field of view of the
  camera. This visual companion has $\Delta$I $\approx$ 7.9
  magnitudes. Further analysis showed that the companion is in fact a
  background star. For more details, see Section \ref{tertiary}.
\label{companion}}
\end{figure}

\subsubsection{Lucky Imaging}
\label{FC:DataAcquisition}
Lucky imaging (LI) involves acquiring large numbers of observations with
a rapid cadence. A subset of these images are then shifted and
stacked in order to produce nearly-diffraction-limited images. \star
was observed over 3 nights separated by 2 years using FastCam
\citep{osc08} on the 1.5 m TCS telescope at Observatorio del Teide
in Spain. The LI frames were acquired on 9 Oct., 2010; 25 Aug., 2011
and 11 Sep., 2012 in the $I$ band and spanning $\sim 21 \times
21~\rm arcsec^2$ on sky. 

The 11 Oct., 2010 observing run had 100,000 frames with 50 msecs per frame,
the 25 Aug., 2011 run had 60,000 frames with 60 msecs per frame, and the
11 Sept., 2012 run had 75,000 frames at 60 msecs per frame.  The data were
processed using a custom IDL software pipeline.  After identifying
corrupted frames due to cosmic rays, electronic glitches, etc., the
remaining frames are bias corrected and flat fielded.

Lucky image selection is applied using a variety of selection
thresholds based on the brightest pixel (BP) method. This method
involves selecting frames for the final combined LI image based on the
BP in that frame. The selected BP must be below a specified brightness
threshold to avoid selecting cosmic rays or other non-speckle
features.  As a further check, the BP must be consistent with the
expected energy distribution from a diffraction speckle under the
assumption of a diffraction-limited PSF.  The BP's of each frame are
then sorted from brightest to faintest, and the best $X$\% are then
shifted and added to generate a final combined image. Several values
of $X$, the LI threshold, are tried until the best combined image is
generated. The combined images using different LI thresholds are shown
for each of the three observing runs in Figure \ref{companion}. The 11
Sept., 2012 run has a better quality image than the other two runs
because of an instrument upgraded to a higher sensitivity CCD. The
best LI threshold was determined to be $50$\%, which is what we used
for the rest of the analysis. This results in a total exposure time
for each of the final combined images from the three runs in order of
date 2500, 1800, and 2250 s respectively.

\begin{figure}[tbp]
\includegraphics[width=\linewidth]{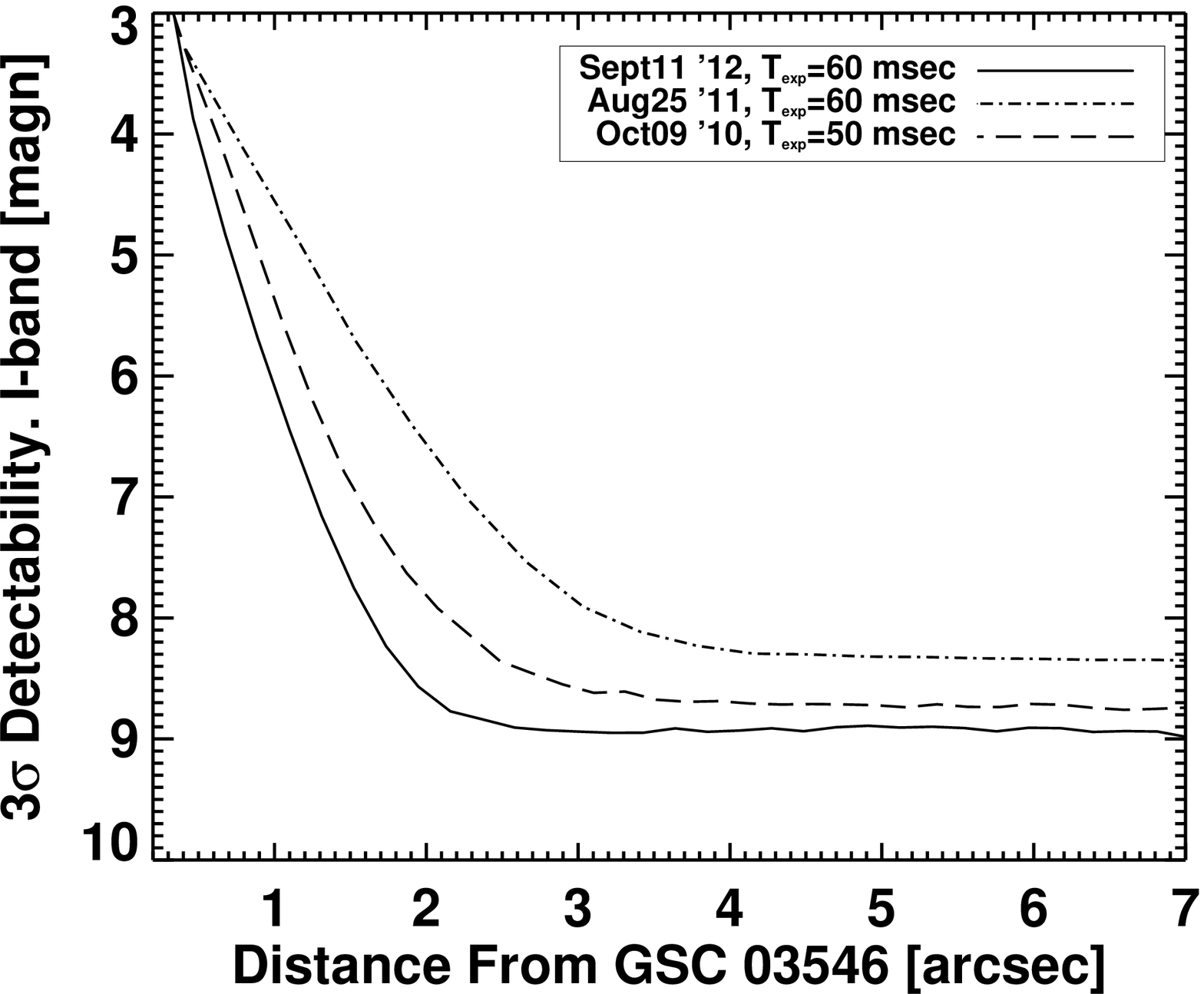}
\includegraphics[width=\linewidth]{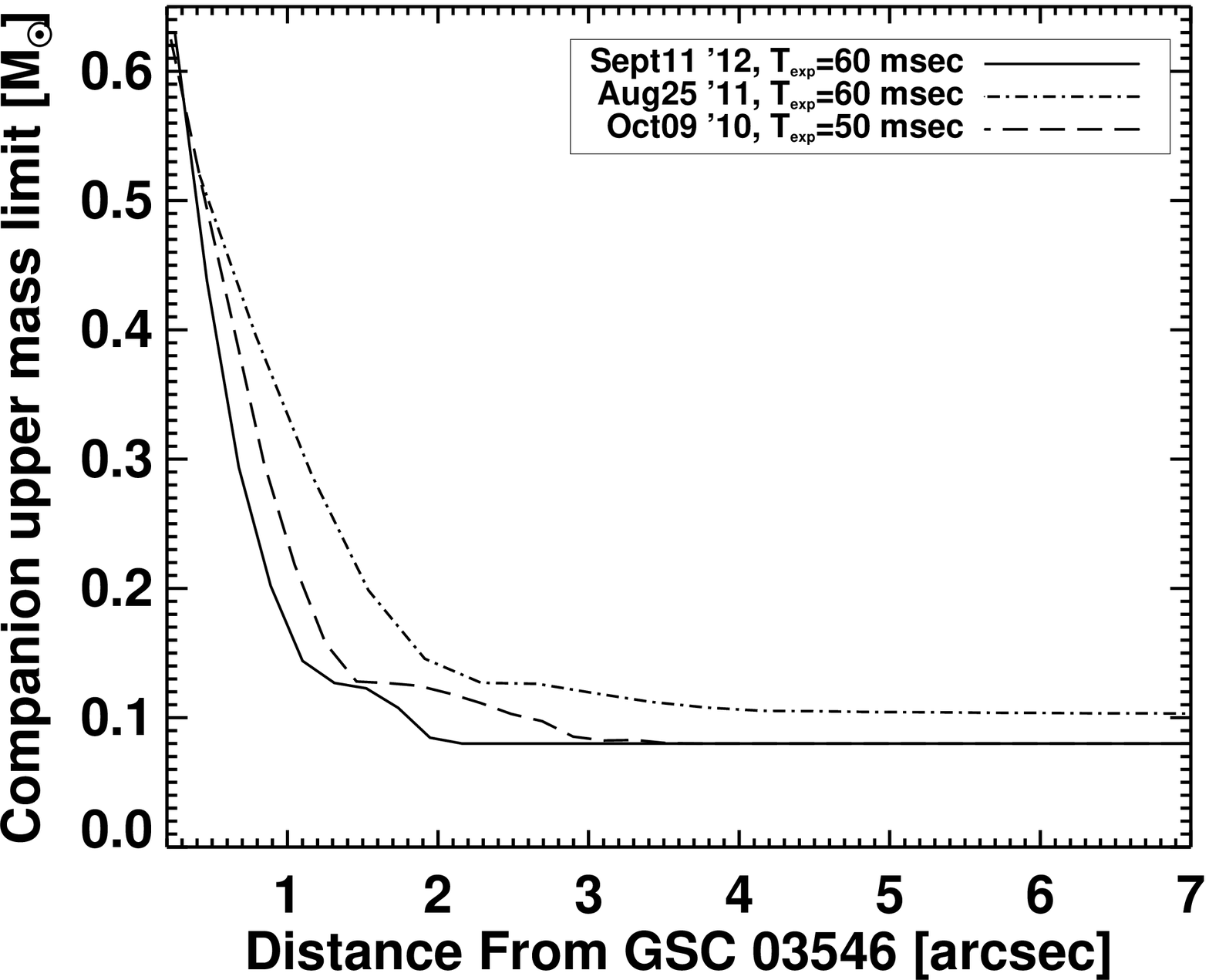}
\caption{Top: 3-$\sigma$ detectability curves for LI from the FastCam
  imager. There was significant variation in the curves over the three
  runs. The best run can detect sources down to $\Delta \rm I = 8.5$
  beyond 2\arcsec. No source was detected within 7\arcsec~of \starnsp. Bottom: Conversion of 3-$\sigma$ detectability
  curves for the three observing nights into mass sensitivities using
  empirical Mass-Luminosity relationships in the literature. See
  Section \ref{tertiary}.
\label{fig:MassConstraint}}
\end{figure}

In Figure \ref{fig:MassConstraint} we show the 3-$\sigma$
detectability curves computed out to 8\arcsec~from \starnsp. We follow
the same procedure as in \citet{fem11} to compute these curves: at a
given angular distance $\rho$ from \star we identify all possible sets
of small boxes of a size larger but comparable to the FWHM of the PSF
(i.e. $5 \times 5$ pixel boxes). Only regions of the image showing
structures easily recognizable as spikes due to diffraction of the
telescope spider and/or artifacts on the read-out of the detector are
ignored. For each of the valid boxes on the arc at angular distance
$\rho$ the standard deviation of the image pixels within the $5\time
5$-pixel boxes is computed. The value assigned to the 3-$\sigma$
detectability curve at $\rho$ is 3 times the mean value from the
standard deviations of all the eligible boxes at $\rho$. This
procedure, using each of the LI \% thresholding values provides a
detectability curve, while the envelope of all the family of curves
for a given night yields the best possible detectability curve to be
extracted from the whole data set.

Although there are no companions detected within 7\arcsec, we do find
a possible companion slightly outside this region at a separation of
7.7\arcsec. This companion is significantly dimmer than \star with a
$\Delta$I $\approx$ 7.9 magnitudes. This object was only detected in
the 11 Oct., 2010 and 11 Sept., 2012 campaigns. The visual companion
was not detected in the 25 Aug., 2012 campaign due to the orientation
of the camera on the sky. The possible companion can be seen in Figure
\ref{companion}. A discussion of the likelihood that this object is
actually a bound companion to \star can be found in Section \ref{tertiary}. 

\section{Results}
\label{results}
\subsection{Host Star Characterization}
\subsubsection{Spectroscopic Analysis}
The ARCES moderate resolution spectrum (Section \ref{apospec}) was
analyzed by two independent analysis pipelines. We refer to these
pipeline results as the ``IAC'' (Instituto de Astrof\'{i}sica de
Canarias) and ``BPG'' (Brazilian Participation Group) results.
Briefly, both methods are based on the principals of Fe I and Fe II
excitation and ionization equilibria. Both techniques employ the 2002
version of the MOOG code \citep{sne73}, but use different line lists,
model atmosphere grids, equivalent width measurements, and convergence
criteria. For a complete description of the analysis process, please
refer to \citet{wis12}.

Our IAC analysis used 204 Fe I lines and 20 Fe II lines, and resulted
in the following parameters: $T_{\rm eff} = 5502 \pm 100$, $\log{g} =
4.21 \pm 0.58$, and [Fe/H] $= +0.31 \pm 0.16$. The BPG
analysis used 61 Fe I lines and 6 Fe II lines resulting in these
parameters: $T_{\rm eff} = 5652 \pm 75$, $\log{g} = 4.46 \pm 0.16$,
and [Fe/H] $= +0.44 \pm 0.10$. Since the values for both
methods are consistent to within 1-$\sigma$ they were combined using a
weighted average as described in \citet{wis12}. This calculation
resulted in the final spectroscopic parameter values of $T_{\rm eff} =
5598 \pm 63$, $\log{g} = 4.44 \pm 0.17$, and [Fe/H] $= +0.40
\pm 0.09$. These values are recorded in Table \ref{hosttab}, and
are used for the remaining analysis.

From these values of $T_{\rm eff}$, $\log{g}$, and [Fe/H]
and their errors, and using the \citet{tor10} relations, including
intrinsic scatter and the uncertainties in the polynomial coefficients
and covariances, we find that the mean and RMS of the mass and radius
of the host are $M_*= 1.11 \pm 0.11~M_\sun$ and $R_*= 1.06 \pm
0.23~R_\sun$ The median and 68 percent confidence intervals are $M_*=
1.11 _{-0.10}^{+0.11}~M_\sun$ and $R_*=1.03_{-0.19}^{+0.26}~R_\sun$.

These spectroscopic values also allows us to transform the detection
curves from the left panel of Figure \ref{fig:MassConstraint} into
upper limits on the mass of a possible stellar companion undetected at
the 3-$\sigma$ level. We start by taking the spectroscopic $T_{\rm
  eff}$ from Table \ref{hosttab}, and using the tables from
\citet{mam10}, we derive absolute V and I-band magnitudes. We can
apply this technique because \star has been identified as a
main-sequence star by the $\log{g}$ measurement. The $M_I$ and the
3-$\sigma$ detectability curves allows the construction of the $M_I$
vs. angular distance, $\rho$ from the central star. This curve provides
an absolute upper I-band limit to any companion, which would also have
to be a main-sequence star. From the $M_V$ vs. $\rho$ curves we can use
empirical Mass Luminosity Relationships (MLR)
\citep{hen99,del00,hen04,xia08,xia10} to derive the upper mass
limit. These constraints are plotted for all three observation nights
in the right panel of Figure \ref{fig:MassConstraint}.

\label{specanal}

\begin{deluxetable*}{lrrc}
  \tabletypesize{\footnotesize}
  \tablewidth{\linewidth}
  \tablecaption{Host Star Properties \star \label{hosttab}}
  \tablehead{
    \colhead{Parameter} 
    & \colhead{Value} 
    & \colhead{1-$\sigma$ Uncertainty} 
    & \colhead{Source} 
  }
\startdata
GSC1.1 Name & 3546-01452 & \nodata &\citet{las08}\\
GSC2.3 Name & N2EA000110 & \nodata &\citet{las08}\\
KIC Name & 11022130 & \nodata &\citet{bro11}\\
2MASS Name & J19113252+4830436 & \nodata &\citet{cut03}\\
$\alpha$ (J2000) [deg] &287.885735 &0.26\arcsec &\citet{las08}\\
$\delta$ (J2000) [deg] & 48.512117 &0.25\arcsec &\citet{las08}\\
$\mu_\alpha$ [$\rm mas~yr^{-1}$] & -28.6& 2.3&\citet{zac12}\\
$\mu_\delta$ [$\rm mas~yr^{-1}$] & -8.7& 1.7&\citet{zac12}\\
$T_{\rm eff}$ [K] &5598 &63 & This work\\
$\log{g}$ [cgs] & 4.44 &0.17 & This work\\
\mbox{[Fe/H]} [dex]& $+0.40$ & 0.09 & This work\\
v$_{mic}$ [$\rm km~s^{-1}$] & 0.38 &0.17 & This work\\
v$_{rot}\ sin\ i$ [$\rm km~s^{-1}$] & $<$9.0 & \nodata & This work\\
v$_{systemic}$ [$\rm km~s^{-1}$] &-13.799&0.034&This work\\
A$_{V}$ [mag]&0.1 &0.1& This work\\
Distance [pc] &219 & 21 & This work\\
M$_{*}$ [M$_{\sun}$] & 1.11 & 0.11 & This work\\
R$_{*}$ [R$_{\sun}$] & 1.06 & 0.23 & This work\\
\\\tableline\tableline\\
galNUV & 18.320 &0.100 & \citet{mor07}\tablenotemark{a}\\
B & 12.593 &0.020 & HAO (This work)\tablenotemark{a}\\
V & 11.799 &0.010 & Derived from HAO\\
R$_C$ & 11.335&0.012 & Derived from HAO\\
I$_C$ & 10.967&0.012 & Derived from HAO\\
SDSS $u$' & 13.483 & 0.020 & HAO (This work) \tablenotemark{a}\\
SDSS $g$' & 12.156&0.012 & HAO (This work)\tablenotemark{a}\\
SDSS $r$' & 11.560&0.010 & HAO (This work)\tablenotemark{a}\\
SDSS $i$' & 11.400&0.010 & HAO (This work)\tablenotemark{a}\\
2MASS J & 10.46 &0.030 & \citet{cut03}\tablenotemark{a}\\
2MASS H & 10.07 &0.030 & \citet{cut03}\tablenotemark{a}\\
2MASS K$_S$ & 10.010 &0.03 & \citet{cut03}\tablenotemark{a}\\
WISE1 & 12.650 & 0.023 & \citet{cut12a}\tablenotemark{a}\\
WISE2 & 13.344 & 0.020 & \citet{cut12a}\tablenotemark{a}\\
WISE3 & 15.181 & 0.032 & \citet{cut12a}\tablenotemark{a}\\
\enddata



\tablenotetext{a}{These passbands were used to generate SED in Figure \ref{hostsed}.}

\end{deluxetable*}

\subsubsection{Photometric Analysis}
We corroborated the spectroscopic results with a series of photometric
tests, including determining the RPM-J statistic from
\citet{col07}. For \starnsp, the RPM-J value was 2.31 and its (J-H) 2MASS
color was 0.387, results consistent with \star being a dwarf. 

\begin{figure}[tbp]
\includegraphics[width=\linewidth]{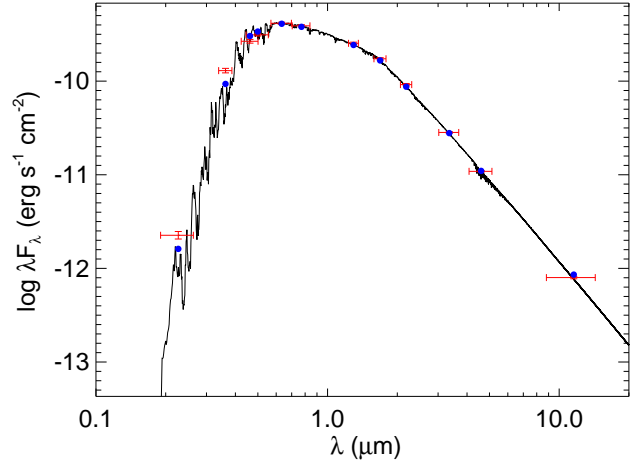}
\caption{The spectral energy diagram (SED) for \starnsp. The black line shows
  the best fit NextGen model compared to the observed flux in the
  photometric passbands from Table \ref{hosttab} (red crosses). The
  horizontal bars are the approximate passband width for each filter,
  while the vertical bar is the error in the flux. The blue circles
  are the expected flux values from the model. The best fit model is
  fixed to the spectroscopically determined $T_{\rm eff}$, $\log{g}$, and
  [Fe/H] and allowed A$_{V}$ to float. The optical and infrared fit
  well, but there is a significant excess of flux in the GALEX NUV passband.
\label{hostsed}}
\end{figure}

We also constructed a spectral energy diagram (SED) of \star using the
\textit{GALEX} NUV filter \citep{mor07}; Johnson B and SDSS
$u'$,$g'$,$r'$,$i'$ observations from HAO; 2MASS J, H, and K
\citep{cut03}; and WISE1, WISE2, WISE3 \citep{cut12a}. A summary of
the photometric values are presented in Table \ref{hosttab}. We used the
NextGen model atmosphere grid \citep{hau99} to construct theoretical
SEDs. These models were fixed by the spectroscopic values for
T$_{Eff}$, $\log{g}$, and [Fe/H] described in Section \ref{specanal};
the reddening was constrained by the \citet{sch98} dust maps in the
direction of ($l$,$b$) = (79\fdg339029, 16\fdg853379) to be less than
A$_V$ = 0.213. The resulting fit to the photometry is shown in Figure
\ref{hostsed}. The best fitting model has a reduced $\chi^2$ of 3.91, a
negligible reddening of A$_V$ = 0.1 $\pm$ 0.1, and a distance of 219
$\pm$ 21 pc. There appears to be a slight UV excess, which may be the
result of modest stellar activity.

\subsubsection{Evolutionary Status and and Galactic Population}
\begin{figure}[tbp]
\plotone{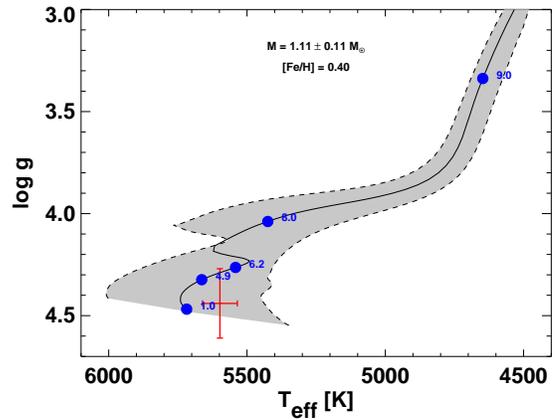}
\caption{The black line represents the best fit Yonsei-Yale
  evolutionary track for a 1.1 M$_{\sun}$ star with a [Fe/H] of +0.40
  \citep{dem04}. The gray area denotes the 1-$\sigma$ deviation from that
  track. Ages along the track are denoted by blue dots for 1.0, 4.7,
  6.4, 8.0, and 9.0 Gyr. The red cross denotes the location of \star on
  this diagram with the spectroscopic uncertainties.
\label{track}}
\end{figure}

Using the spectroscopic host star parameters and the mass and radius
derived from the \citet{tor10} relations, \star can be placed onto an
evolutionary track. We use the Yonsei-Yale (``Y2'') model tracks from
\citet{dem04} (and references therein), and select the track
corresponding to 1.11 $M_{\sun}$ with a [Fe/H] $= +0.40$. Figure
\ref{track} shows this track with stellar ages marked as blue dots on
the track. The dashed lines show tracks for the same metallicity, but
for $\pm 0.11 M_{\sun}$, which is the 1-$\sigma$ uncertainty on our
mass estimate. The gray area shows the region that this family of
tracks occupies. The red point shows \star with 1-$\sigma$ error bars. From
the error bars in $\log{g}$ and $T_{\rm eff}$ in Figure \ref{track} we
can constrain the age of \star to being less than approximately 6 Gyr old.

We can determine the Galactic population membership of \star by using
the absolute systemic RV=$-13.799\pm 0.128$~$\rm km~s^{-1}$ (where the
$0.128$~$\rm km~s^{-1}$ is a conservative estimate of the systematic
error from the comparison with RV standard stars), the proper motions
from UCAC4 \citep{zac12} ($\mu = -28.6 \pm 2.3, -8.7 \pm 1.7$~$\rm
mas~yr^{-1}$), and the distance from Table \ref{hosttab} ($219 \pm 21$
pc). We find $(U,V,W)= (24.0 \pm 2.5, -10.1 \pm 1.3, 25.6 \pm
3.1)$~$\rm km~s^{-1}$.  This velocity is consistent with membership in
the thin disk according to the criteria of \citet{ben03}.  The space
motion velocities were determined using a modification of the IDL
routine {\tt GAL\_UVW}, which is ultimately based on the method
\citet{joh87}. We adopt the correction for the Sun's motion with
respect to the Local Standard of Rest from \citet{cos11}, and choose a
right-handed coordinate system such that positive $U$ is toward the
Galactic Center.

\subsection{Companion Orbital Analysis}
The orbital parameters of \bd were derived using the radial velocity
data from both the MARVELS and the SARG spectrographs. The orbital fit
was created using the EXOFAST\footnote{http://astroutils.astronomy.ohio-state.edu/exofast/} code \citep{eas13}, which allowed us to
determination of the usual Keplerian orbital elements shown in Table \ref{orbittab}. 

The first step in the process was to perform independent fits to both
the MARVELS and SARG data points. Both data sets had error bars that
were too large based on the $\chi^{2}$ of their fits, so we re-scaled
them to force the probability that the $\chi^{2}$ is greater than or
equal to the measured $\chi^2$, i.e., P($\chi^{2}$), =
0.5. The MARVELS error bars were scaled by a factor of 0.579, and
the SARG error bars were scaled by a factor of 0.222. The error
bars in Table \ref{rvtab} include these scaling factors.

\begin{figure}[tbp]
\plotone{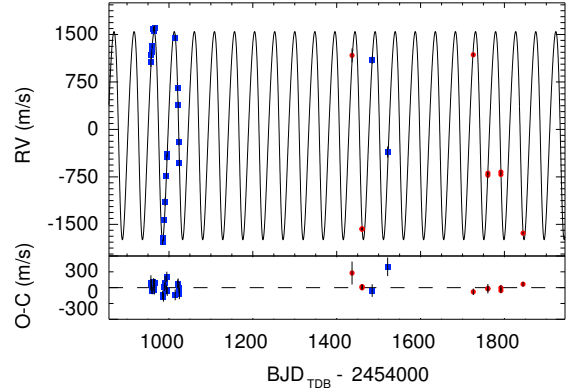}
\caption{All the radial velocity points from both the MARVELS Survey
  (blue squares) and TNG/SARG follow-up spectroscopy (red
  circles). The final best fit for a $M_b \sin i =$ 31.7 $\pm$ 2.0 $M_{Jup}$ companion to \star (Table \ref{orbittab}) is shown with
  the associated O-C diagram. The observation data points start at BJD =
  2454956.916 (5 May, 2009). There is a slight residual linear slope
  (\mbox{0.180 $\rm m~s^{-1}~day^{-1}$}) after removing the companion orbit.
\label{unfolded}
}
\end{figure}

\begin{figure}[tbp]
\plotone{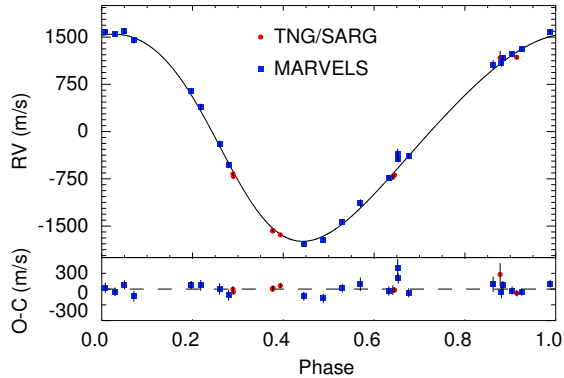}
\caption{The phased final RV curve including both the MARVELS Survey
  (blue squares) and the TNG/SARG observations (red circles). The final
  best fit for a $M_b \sin i =$ 31.7 $\pm$ 2.0 $M_{Jup}$ companion to \star (Table \ref{orbittab}), is shown with the associated O-C
  diagram.
\label{folded}
}
\end{figure}

Once the error bars were scaled, the two data sets were combined. The
final combined radial velocity curve was then run through EXOFAST
which is a Markov Chain Monte Carlo (MCMC) based IDL software
suite. The basic algorithms of this software follow the work of
\citet{for06}. The results of the combined fit are listed in Table 
\ref{orbittab}, and the best-fit model is shown in Figures \ref{unfolded} and \ref{folded}.

The radial velocity measurements from both MARVELS and SARG are
relative radial velocities. As part of the fitting process, EXOFAST
fits and subtracts arbitrary zero points for each data set
simultaneously. These zero points are recorded in Table
\ref{orbittab}. These offsets are based on arbitrary zero-points, and
should not be confused with the true systemic velocity of the host
system $-13.799\pm 0.128~\rm km~s^{-1}$.

The systemic velocity of \star was derived by taking the orbital
solution from EXOFAST for the combined relative radial velocity data
set and applying it to the SARG/TNG absolute velocity data set (Table
\ref{absrvtab}.). This calculation was performed by using the MPFIT
Levenberg-Marquardt non-linear least squares fitting algorithm in IDL
\citep{mar09}. A Keplerian model was fit to the SARG/TNG radial
velocity points, holding $P,~e,~\omega,~K$, and $T_P$ fixed while
allowing $\gamma$ to float; the results are presented in \mbox{Table
 \ref{hosttab}}.
\label{rvanalysis}

\newcommand{\bjdtdb}{\ensuremath{\rm {BJD_{TDB}}}}
\newcommand{\feh}{\ensuremath{\left[{\rm Fe}/{\rm H}\right]}}
\newcommand{\teff}{\ensuremath{T_{\rm eff}}}
\newcommand{\ecosw}{\ensuremath{e\cos{\omega_*}}}
\newcommand{\esinw}{\ensuremath{e\sin{\omega_*}}}
\newcommand{\msun}{\ensuremath{\,{\rm M}_\Sun}}
\newcommand{\rsun}{\ensuremath{\,{\rm R}_\Sun}}
\newcommand{\lsun}{\ensuremath{\,{\rm L}_\Sun}}
\newcommand{\mj}{\ensuremath{\,{\rm M}_{\rm J}}}
\newcommand{\rj}{\ensuremath{\,{\rm R}_{\rm J}}}
\newcommand{\fave}{\langle F \rangle}
\newcommand{\fluxcgs}{10$^9$ erg s$^{-1}$ cm$^{-2}$}
\begin{deluxetable}{lcc}
  \tablecaption{Properties for \bd \label{orbittab}}
  \tablehead{
    \colhead{~~~Parameter} 
    & \colhead{Units} 
    & \colhead{Value}
  }
\startdata
                 $T_C$\dotfill &\bjdtdb \ - 2450000\dotfill & $5023.377_{-0.099}^{+0.10}$\smallskip\\
                         $P$\dotfill &Period (days)\dotfill & $47.8929_{-0.0062}^{+0.0063}$\smallskip\\
                          $e$\dotfill &eccentricity\dotfill & $0.1442_{-0.0073}^{+0.0078}$\smallskip\\
 $\omega$\dotfill &argument of periastron (radians)\dotfill & $1.998_{-0.061}^{+0.064}$\smallskip\\
               $K$\dotfill &RV semi-amplitude (m/s)\dotfill & $1644_{-13}^{+12}$\smallskip\\
$\gamma_{TNG}$\dotfill &TNG Systemic velocity (m/s)\dotfill & $724\pm15$\smallskip\\
                $dv/dt$\dotfill &RV slope (m/s/day)\dotfill & $0.180\pm0.053$\smallskip\\
$\gamma_{APO}$\dotfill &APO Systemic velocity (m/s)\dotfill & $1058_{-16}^{+17}$\smallskip\\
                          $e \cos(\omega)$\dotfill & \dotfill & $-0.0597_{-0.0070}^{+0.0071}$\smallskip\\
                          $e \sin(\omega)$\dotfill & \dotfill & $0.1312_{-0.0092}^{+0.0093}$\smallskip\\
                 $T_P$\dotfill &\bjdtdb \ - 2450000\dotfill & $5025.81_{-0.44}^{+0.46}$\smallskip\\
 $M_b \sin i$\dotfill & Minimum Mass ($M_{Jup}$)\dotfill & $31.7_{-2.0}^{+2.0}$\\
\enddata
\end{deluxetable}

\subsection{Companion Mass}
\subsubsection{Mass Functions of Secondary}
The mass function (${\cal M}_b$) is the only property of a
single-lined radial velocity companion that we can derive that is
independent of the properties of the primary.  Using the MCMC chain
from the joint RV fit, we can derive ${\cal M}_b$ of the companion,
\begin{eqnarray}
{\cal M}_b &\equiv& \frac{(M_b \sin i)^3}{(M_*+M_b)^2} \nonumber\\
 &=& K^3 (1-e^2)^{3/2} \frac{P}{2\pi G}\\ 
 &=& (2.136 \pm 0.049) \times 10^{-5} M_\sun, \nonumber
\label{eqn:mf}
\end{eqnarray}
where the uncertainty is essentially dominated by the uncertainty in $K$,
such that $\sigma_{{\cal M}_b}/{\cal M}_b \sim  3(\sigma_{K}/{K}) = 3\times 0.8\% \sim  2.4\%$.  
\subsubsection{Minimum mass and mass ratio}
In order to determine the mass or mass ratio of the secondary, we 
must estimate the mass of the primary, as well as the inclination
of the secondary.  

To estimate the mass and radius of the primary, we proceed as follows.
For each link in the MCMC chain from the joint RV fit, we draw a value
of $T_{\rm eff}$, $\log{g}$, and [Fe/H] for the primary from Gaussian
distributions, with means and dispersions given in \mbox{Table
  \ref{hosttab}}. We then use the \citet{tor10} relations to estimate
the mass $M_*$ and radius $R_*$ of the primary, including the
intrinsic scatter in these relations.

The minimum mass (i.e., $M_b$ if $\sin i=1$) and minimum mass ratio of
the secondary are:

\begin{eqnarray}
M_{b,\rm min}&=& 31.7 \pm 2.0~M_{Jup} \nonumber\\
          &=& 0.0303 \pm 0.0020~M_\sun \\
q_{min} &=& 0.02733 \pm 0.00092 \nonumber
\label{eqn:m2min}
\end{eqnarray}
The uncertainties in these estimates are almost entirely explained by
the uncertainties in the mass of the primary: $\sigma_{M_b}/M_b \sim
(2/3)(\sigma_{M_*}/M_*) = (2/3)\times 9.7\% \sim 6.5\%$, close to the uncertainty
in $M_{b,\rm min}$ above ($6.4\%$), and $\sigma_q/q \sim (1/3)(\sigma_{M_*}/M_*) \sim (1/3)\times 9.7\%
\sim 3.2\%$, close to the uncertainty in $q$ ($3.4\%$).  

\subsubsection{\emph{A posteriori} distributions of the true mass}
\label{truemass}
The \emph{a posteriori} distribution of the true mass of the companions given
our measurements depends on our prior distribution for the mass of the
companion, which is roughly equivalent our prior on the mass ratio.

If we assume a prior that is uniform in the logarithm of the true mass
of the companion, then the posterior distribution of $\cos i$ will be
uniform.  More generally, for other priors, $\cos i$ is not uniformly
distributed.  We adopt priors of the form:
\begin{equation}
\frac{dN}{dq} \propto q^{\alpha}
\label{eqn:qprior}
\end{equation}
where $q$ is the mass ratio between the companion and the primary, and
$\alpha=-1$ for the uniform logarithmic prior discussed above.  To include
this prior, for each link the MCMC chain we draw a value of $\cos i$
from a uniform distribution, but then weight the resulting values of
the derived parameters for that link (i.e., the companion mass $m$) by
$q^{\alpha+1}$.

For $\alpha>0$, the \emph{a posteriori} distribution does not coverage.
However, we can rule out equal mass ratio companions by the lack of a
second set of spectral lines.  We therefore assume that the mass ratio
cannot be greater than unity; i.e., we give zero weight to
inclinations such that $q>1$.  In doing so, we implicitly assume that
the companion is not a stellar remnant.

\begin{figure}[tbp]
\plotone{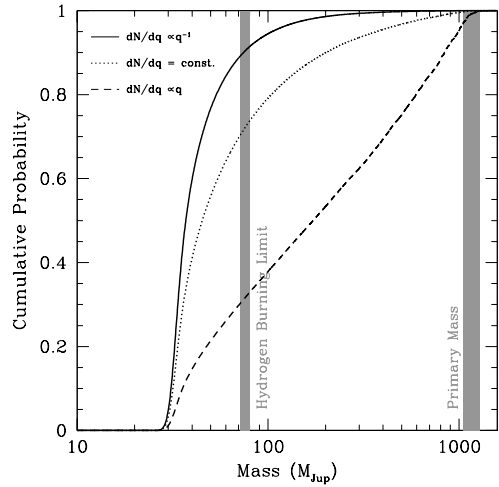}
\caption{ The cumulative \emph{a posteriori} probability of the true
  mass of the secondary, for three different priors on the companion
  mass ratio, $dN/dq \propto q^{\alpha}$, with $\alpha=-1$ (solid),
  $\alpha=0$ (dotted) and $\alpha=1$ (dashed).  Also indicated are the
  mass of the primary with uncertainty and the hydrogen burning limit
  both denoted by shading. For $\alpha=-1$ and $\alpha=0$, the inferred
  median masses are $\sim 37~M_{Jup}$ and $\sim 45~M_{Jup}$,
  respectively.  For these priors, we conclude that the companion has
  a mass below the hydrogen burning limit and thus is a \emph{bona
    fide} brown dwarf at $90\%$ and $72\%$ confidence, respectively.
  For $\alpha=1$, the median mass is $172~M_{Jup}$ or $\sim
  0.16~M_\sun$, firmly within the stellar regime. However, it is
  unlikely that $\alpha > 0$ in this mass regime.
\label{fig:mass}
}
\end{figure}

Figures \ref{fig:mass} shows the resulting cumulative \emph{a posteriori}
distributions of the true mass of the companion, for $\alpha=-1$
(uniform logarithmic prior on $q$), $\alpha=0$ (uniform linear prior),
and $\alpha=1$.  For $\alpha=-1$ and $\alpha=0$, the inferred median
masses are $\sim 37~M_{Jup}$ and $\sim 45~M_{Jup}$, respectively.  For these
priors, we conclude that the companion has a mass below the hydrogen
burning limit and thus is a \emph{bona fide} brown dwarf at $90\%$ and $72\%$
confidence, respectively.  For $\alpha=1$, the median mass is
$172~M_{Jup}$ or $\sim 0.16~M_\sun$, firmly within the stellar regime.
Indeed, for this prior, there is only a $\sim 32\%$ probability that
the companion is a brown dwarf.  However, this conclusion is sensitive
to the precise form for our constraint that $q\le 1$. With a more
careful analysis it may be possible to rule out a wider range of
companion masses based on the lack of evidence for a second set of
spectral lines.  Furthermore, there is little evidence that the mass
function of companions to solar type stars is rising as steeply as
$\alpha=1$ for minimum masses around $30~M_{Jup}$ \citep{gre06}.

We conclude that the companion to \star is most likely a brown
dwarf.  However, we cannot definitively exclude that it is, in
fact, a low-mass stellar companion seen at low inclination.

\subsection{Time-Series Photometric Analysis}
\subsubsection{Summary of Datasets}
The SuperWASP photometric dataset for \star consists of 8823
points spanning just over four years from HJD$'=3128$ to 4690 (where
HJD$' \equiv \rm HJD - 2450000$).  The full,
detrended SuperWASP dataset has a relatively high weighted RMS of $1.8\%$
and exhibits evidence for systematics.  The distribution of residuals
from the weighted mean is asymmetric and non-Gaussian, showing
long tails containing a much larger number of $>3$-$\sigma$ outliers than
would be expected for a normally-distributed population.  We therefore
cleaned the SuperWASP data as follows.  We first add a trial systematic
fractional error in quadrature to the photon noise uncertainties
$\sigma_{sys}$.  We then compute the error-weighted mean flux, and
determine and reject the largest error-normalized outlier from the mean
flux.  We then recompute the mean flux, and scale the uncertainties by
a constant factor $r$ to force $\chi^2/{\rm dof}=1$. We iterate this
procedure until no more $>4$-$\sigma$ outliers remain.  We then repeat
the entire procedure to determine the value of $\sigma_{sys}$ that
results in a distribution of normalized residuals that has the
smallest RMS from the Gaussian expectation.  Although 4-$\sigma$ is a
slightly larger deviation than we would expect based on the final
number of points, we adopt this conservative threshold in order to
avoid removing a potential transit signal.  We adopt $r=0.60$ and
$\sigma_{sys}=0.011$.  The cleaned light curve has 8635 data points
with an RMS of $1.5\%$ and $\chi^2/{\rm dof}=1$ (by design).

The Allengheny photometric dataset consists of 3990 points spanning
roughly a year and a half from HJD$'$=5341 to 5884.  The weighted RMS
of the raw light curve is $0.63\%$.  There is mild evidence for
systematics errors in this dataset, and thus we repeat the identical
procedure as with the SuperWASP data. We choose a more conservative
6-$\sigma$ cut, given the better precision of the Allengheny
observations, to prevent us from removing real transit-like
variability.  We adopt $r=1.04$ and $\sigma_{sys}=0.0028$, with a
final RMS of $0.4\%$ from 3983 data points.

While continuous variability will not be removed by our procedure,
variability in the form of a small number of highly-discrepant point
will be masked.  This possibility is particularly relevant for transit
signatures.  Given the typical uncertainties of $\sim 0.4\%$ for that
data set, we do not expect our cleaning procedure to remove outliers
that differ by less than $\sim 2.4\%$ from the mean, roughly
corresponding to the typical depth for a transit of a $\sim
1.5~R_{Jup}$ companion given the primary radius of $\sim
1.03~R_\sun$.

\begin{figure}[tbp]
\plotone{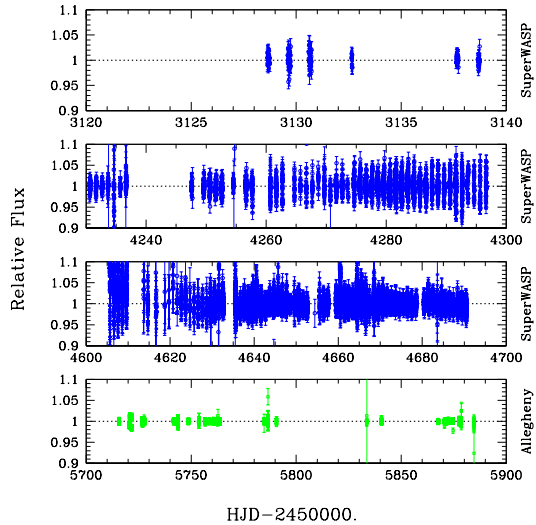}
\caption{Cleaned broad band relative photometry of \star
  from SuperWASP (blue) and Allegheny (green) as a function of date.  
\label{fig:lc}}
\end{figure}

Finally, we combine all the relative photometry after normalizing each
individual data set by its mean weighted flux.  Figure \ref{fig:lc}
shows the combined data set, which consists of 12618 data points
spanning roughly 7.6 years from HJD$'=3128$ to 5884.  The weighted RMS
is $0.66\%$.  The resulting light curve is constant to within the
uncertainties over the entire time span.  Because the individual data
sets sample disjoint times and are not relative common set of
reference stars, our procedure of normalizing each data set by its
mean flux will remove variations on the longest timescales.

\subsubsection{Search for Variability}

\begin{figure}[tbp]
\plotone{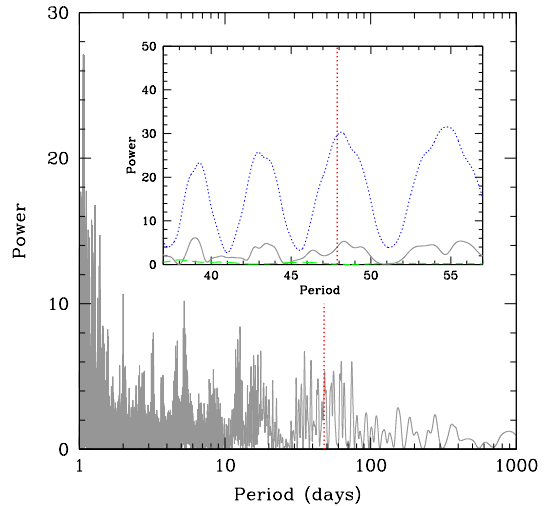}
\caption{Lomb-Scargle periodogram of the combined photometric data.
  While a large number of peaks are visible, we do not regard any of
  them as significant.  In particular, while there is a local peak at
  the period of the secondary (vertical red dotted line), this peak
  arises from the SuperWASP data alone, and is not confirmed by the
  Allegheny data.  This result is demonstrated in the inset, which
  shows the combined periodogram near the period of the companion
  (grey solid), the periodogram of the SuperWASP data alone (blue dotted),
  and Allegheny alone (blue long dashed).
\label{fig:powbin}}
\end{figure}

We ran a Lomb-Scargle periodogram on the full photometric dataset,
testing periods between $1-10^3$ days. The resulting periodogram shown
in Figure \ref{fig:powbin} does not exhibit any strong peaks.  The
maximum fitted amplitude over this range is $\sim 0.1\%$.  The inset
presents the periodogram of the combined data set for periods within
$\sim 10$ days of the period of the companion.  The individual
periodograms for the SuperWASP and Allengheny data are also shown.  While
the SuperWASP data display a local peak at the period of the companion,
this feature is not corroborated by the Allengheny data.  Although
this diffrence could in principle arise from real variability that is
present in the SuperWASP data, but not in the Allegheny data (i.e.,
evolving spots), given the higher precision of the Allengheny data, we
believe it is more likely that the peak in the SuperWASP periodogram is due to
low-level systematics in the form of residual correlations on a range
of time scales.

\begin{figure}[tbp]
\plotone{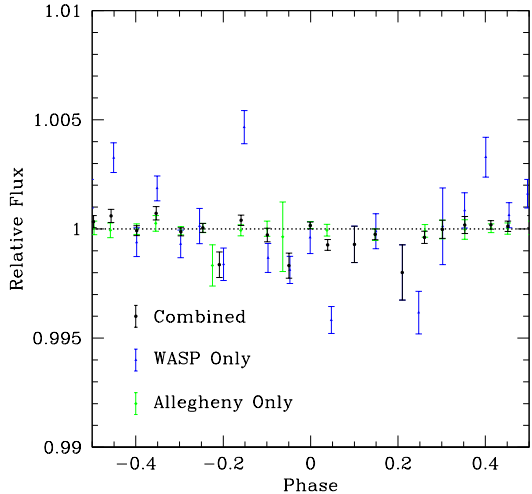}
\caption{
Relative photometry folded at the period of secondary, and binned 0.05
in phase.  Phase zero corresponds to the expected time of conjunction
(and so of transits for the approprate inclinations).  Black points
are the combined data, blue are SuperWASP, and green are
Allegheny. \label{fig:binall}}
\end{figure}

Figure \ref{fig:binall} shows the combined light curve, folded at the
median period and time of conjuction of the companion, ($P=47.8929\pm
0.0063$ and $T_C=5023.377$), and binned 0.05 in phase. The weighted
RMS of the binned light curve is $\sim 0.041\%$.  Although the
variations are larger than expected from a constant light curve based
on the uncertainties $(\chi^2/{\rm dof} \sim 2.6$), we again suspect
that these are due to systematic errors in the relative photometry. In
particular, the folded, binned Allegheny light curve shows a somewhat
lower RMS of $\sim 0.016\%$, and is more consistent with a constant
flux with a $\chi^2/{\rm dof} = 0.4$.

We conclude that there is no strong evidence for variability of
\star on any time scale we probe.  We can robustly
constrain the amplitude of any persistent periodic variability to be
less than $\la 0.1\%$ over timescales of $\la 3$ years, and we can
constrain the amplitude of persistent photometric variability at the
period of the companion to $\la 0.02\%$.

\subsubsection{Excluding Transits of the Secondary}

For a uniform distribution in $\cos{i}$ (corresponding to a prior that
is uniform in the logarithm of the mass), the secondary transit
probability is $\sim 1.8\%$.  The duration for a central transit is
$\sim R_*P/(\pi a) \sim 6.6$ hours, and the depth for Jupiter-radius
body is $\delta\sim (R_b/R_*)^2 = 1.0\% (R_b/R_{Jup})^2$, where $R_b$ is the
radius of the companion, and we have adopted the median value for the
primary radius of $R_*=1.03~R_\sun$.  Thus, the expected S/N of a
transit in the combined photometric dataset assuming uniform phase
coverage is
\begin{equation}
{S/N} \sim N^{1/2} \left(\frac{R_*}{\pi a}\right)^{1/2}
\frac{\delta}{\sigma}\sim 13 \left(\frac{R_b}{R_{Jup}}\right)^2
\label{eqn:transitsnr}
\end{equation}
where $N=12618$ is the number of data points, and $\sigma \sim 0.66\%$
is the RMS of the combined light curve.  Thus we would expect to be
able to robustly detect or exclude transits from companions with radii
as as small as $\sim 0.9~R_{Jup}$ at 10-$\sigma$ if the transit phase is
well covered by the photometric data.  However, given the long period
of the companion, uniform phase coverage is not necessarily a good
approximation.

Therefore, in order to account for the actual sampling of the
photometric data, we perform a quantitative search for transit signals
using a modified Monte Carlo method.  This method is similar to that
described in \citet{fle12} we briefly review the details here.  We
start with the distributions of the relevant radial velocity fit
parameters: the period $P$, semiamplitude $K$, eccentricity $e$, and
time of conjuction $T_C$.  These are obtained from the MCMC chain
determined by fitting of the joint radial velocity data set as
described in \ref{rvanalysis}.  For each link in this MCMC chain, we
also draw values for $T_{\rm eff}$, $\log{g}$, and [Fe/H]
for the primary from Gaussian distributions, with means and
dispersions given in Table \ref{hosttab}.  We then use the
\citet{tor10} relations to estimate the mass $M_*$ and radius $R_*$ of
the primary, including the intrinsic scatter in these relations.  We
then draw a value of $\cos i$ from a uniform distribution, assuming a
prior on the companion mass $M_b$ that is uniform in $\log{M_b}$.  The
values of $P$, $K$, $e$, $M_*$, $R_*$ and $i$ are then used to
determine the secondary mass $M_b$, semimajor axis $a$, and impact
parameter of the secondary orbit $b \equiv a\cos{i}/R_*$.  From the
impact parameter, we determine if the companion transits ($b \le 1$),
and, if so, we determine the properties of the light curve using the
routines of \citet{man02} for a given companion radius $R_b$.  We assume
quadratic limb darkening, adopting coefficients appropriate for the
$R$ band from \citet{cla03}, assuming solar metallicity and the
central values of $T_{\rm eff}$ and $\log{g}$ listed in Table
\ref{hosttab}.  Finally, we fit the predicted transit light curve to
the combined photometric data, and compute the $\Delta \chi^2$ between
the constant flux fit and the transit model.  We perform this
calculation for every step in the Markov Chain, and finally repeat
this procedure for a range of companion radii.

We find a best-fit transit model has $\Delta \chi^2=-9.0$ relative to
a constant fit.  We do not consider this result to be significant, as
we find a larger improvement $\Delta \chi^2=-28.1$ when we consider
signals with the same ephemeris and shape as transits, but
corresponding to positive deviations (i.e., ``anti-transits'', see
\citet{bur06}).  Although formally statistically significant, both the
transit and anti-transit signals are likely caused by residual
systematics in the cleaned photometric data set, and thus we
conclude there is no evidence for a transit signal in the photometric
data.

\begin{figure}[tbp]
\plotone{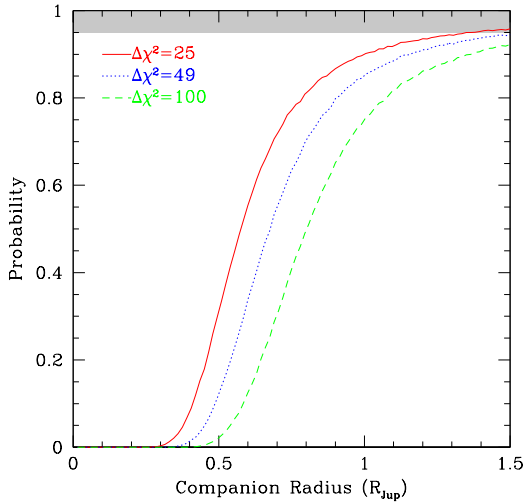}
\caption{Probability that transits of a companion are excluded at
  levels of $\Delta\chi^2=25, 49, 100$ based on the analysis of the
  combined SuperWASP and Allegheny photometric data sets, as a function of
  the radius of the companion. The gray region represents the 95 percent
  confidence level. Only the $\Delta\chi^2=25$ model crosses into the
  gray region, thus we cannot rule out a possible transiting companion
  even with an unphysically large radii $R_b\ga 1.5~R_{Jup}$ at the 95\%
  level. \label{fig:exctrans}}
\end{figure}

This procedure can be used to determine the confidence with which
we can rule out transits of a companion with a given radius.
Specifically, the confidence with which we can exclude a companion
with a given $R_b$ is simply the fraction of the steps in the Markov
Chain where the companion transits, which produces a transit with a
$\Delta\chi^2$ relative to the constant fit that is greater than some
value of $\Delta\chi^2$.  We consider three different thresholds of
$\Delta\chi^2= 25$, 49, and 100.  The resulting cumulative probability
distributions are shown in Figure \ref{fig:exctrans}.  Given the
values of $\Delta\chi^2$ found for anti-transits, we conservatively
consider thresholds of $\Delta\chi^2 \ga 49$ to be robust.  For this
value, we can exclude companions with $R_b \ga 0.7~R_{Jup}$ with 50\%
confidence (i.e., for 50\% of the trials), and companions with $R_b \ga
1.2~R_J$ with 90\% confidence.  However, there is a long tail toward
large companion radii, arising from the conflation of the long period
of the companion, uncertainties in the emphemeris, and imperfect phase
coverage.  Therefore, we are unable to exclude transits at $>95\%$
confidence even for $R_b\ga 1.5$.  For the minimum companion mass of
$\sim 32~M_{Jup}$, \citet{bar03} predict radii of $\sim 0.9-1~R_{Jup}$ for
ages between 0.5 and 5 Gyr.  Therefore, we cannot definitively exclude
the possibility that the companion transits.

As discussed above, our 6-$\sigma$ clipping procedure would in
principle remove $\ga 2.4\%$ transit signatures from the Allegheny
dataset arising from a large $R_b\ga 1.5~R_{Jup}$ companion.  We
therefore repeated the transit search on the Allegheny dataset with
10-$\sigma$ clipping, but again find no significant transits.

\subsection{LI Tertiary Companion}
\label{tertiary}
As discussed in Section \ref{FC:DataAcquisition} a possible tertiary
companion was detected at 7.7\arcsec~from \star (see Figure
\ref{companion}). The possible tertiary companion has $\Delta$I
$\approx$ 7.9 magnitudes, and at the measured distance to the host
star in Table \ref{hosttab} places the companion at a projected
distance of 1686 AU from the host star.

This possible tertiary companion can be found in both the 2MASS
\citep{cut03} and \emph{Kepler} Input Catalog
\citep[KIC;][]{bro11}. The KIC ids of the primary and possible
tertiary are 11022130 and 11022139 respectively. The first blow to the
hypothesis that this is a bound system is found in the KIC itself. The
KIC gives the following stellar parameter estimates for the possible
tertiary $T_{\rm eff} = 5919$, $\log{g} = 4.361$, and $A_V =
0.515$. Comparing these values to the spectroscopic values determined
for \star in Table \ref{hosttab} it is clear that these two objects
cannot be bound, and have a $\Delta$I of $\approx$ 7.9 magnitudes;
because two G dwarfs should have approximately the same
magnitude. However, the tertiary is faint (J = 16.272),
and the KIC spectroscopic determinations are photometric in nature, so
we considered the possibility that the KIC parameters were incorrect.

\star has a measured proper motion from the UCAC4 catalog
\citep{zac12} of $\mu_\alpha$ = -28.6 $\pm~2.3~\rm mas~yr^{-1}$ and
$\mu_\delta$ = -8.7 $\pm~1.7~\rm mas~yr^{-1}$ (see Table
\ref{hosttab}). This motion is significant given the plate scale of
the LI images $\approx$ 42.56 mas $\rm pixel^{-1}$, which, given our 2
year baseline for observations, should be sufficient to test whether
this is a bound system. Table \ref{postab} gives the difference in
position between the \star and the possible tertiary in RAcos(DEC) and
DEC for both LI observations. The two points are distinct to the
2-$\sigma$ level. We also have observations of both sources in 2MASS,
although the astrometric position measurement is significantly worse,
the 12 year baseline between 2MASS and the 2012 observation
compensates for the lower accuracy. These points are also distinct to
the 2-$\sigma$ level. Given both the astrometric and photometric data,
the hypothesis that this is a bound system is very unlikely.

\begin{deluxetable}{cccccc}
  \tabletypesize{\footnotesize}
  \tablewidth{0pt}
  \tablecaption{Difference in Position of \star versus Possible Tertiary\label{postab}}
  \tablehead{
    \colhead{Year Observed} 
    & \colhead{$\Delta\alpha\cos(\delta)$} 
    & \colhead{Error} 
    & \colhead{$\Delta\delta$}
    & \colhead{Error}
    & \colhead{Source}\\
    \colhead{\nodata}
    & \colhead{(arcsec)}
    & \colhead{(arcsec)}
    & \colhead{(arcsec)}
    & \colhead{(arcsec)}
    & \colhead{\nodata}
  }
\startdata
2012& 7.232& 0.049& -3.026& 0.043 & Lucky\\
2010& 7.270& 0.023 &-3.163& 0.018& Lucky\\
2000& 6.86&0.11& -3.39& 0.12&2MASS\\
\enddata
\end{deluxetable}

\section{Discussion}
Based on the derived spectroscopic parameters for the host star and
the results discussed in Section \ref{truemass}, it is highly likely
that \bd is a BD companion to \star assuming a logarithmic or linear
prior on the distribution of mass ratios. Working with the minimum
mass for \bd of 31.7 $\pm$ 2.0 $M_{Jup}$, we can compare it to current
distribution of masses in the literature. \citet{gre06} investigated
the mass function of both stellar and planetary companions and found
that they both decreased by a substantial fraction as they approached
the minimum of the BD desert. In the case of the stellar mass
function, it dropped by two orders of magnitude from the 1 $M_{\sun}$
to the middle of the BD desert. On the planetary mass side, it rises
one order of magnitude as one progress away from the BD desert and
towards lower mass planets.

\bd is interesting because its minimum mass, 31.7 $\pm$ 2.0 $M_{Jup}$,
lies at the very bottom of the two distributions. \citet{gre06} define
their mass functions as number of companions per unit interval in log
mass, and find the minimum of the distributions to be at
$31^{+25}_{-18}M_{Jup}$. In order to better understand this minimum,it
is necessary to increase the number of objects (or place tighter
constraints by non-detections) in this region of the mass function, and
\bd takes us one step closer to that goal.

\begin{figure}[tbp]
\plotone{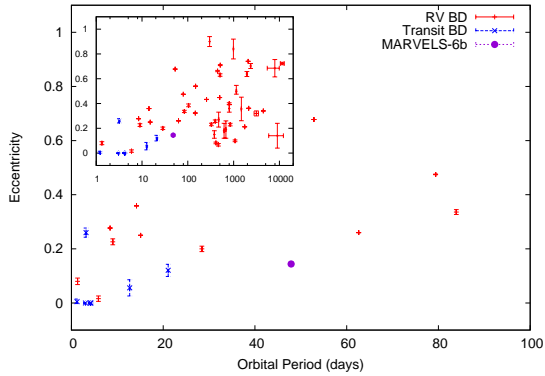}
\caption{Orbital period and eccentricity of \bd compared to a subset
  of literature brown dwarfs found through transit or radial velocity
  that occupy the period range of 0 to 100 days. The literature brown
  dwarfs are from the catalog of \citet{ma13b}. \bd occupies an empty
  part of orbital period / eccentricty plane. Inset: The full brown
  dwarf catalog extending out to much larger periods.
\label{perecc}
}
\end{figure}

\bd stands out in respect to other brown dwarfs in two significant
ways. First, it has a relatively low eccentricity, 0.1442, given its
moderately long period of $\approx$ 47 days. To place this point in
perspective, \bd is plotted against a catalog of BDs from
\citet{ma13b} in Figure \ref{perecc}. Two aspects of this plot are
particularly interesting: First, \bd is in an underdense region in
period space. In the inset of Figure \ref{perecc} there is a clear
rise in the envelope of eccentricity as one moves to larger periods
until one reaches roughly 200 days. \bd is either on or below the
bottom edge of this envelope. There is the possibility that this
effect is due to observational bias; for low signal to noise ratios
(SNR), one often measures a non-zero eccentricity even when the true
eccentricity is zero. For a given radial velocity precision and
companion mass, longer periods will have lower SNR (since the
semi-amplitude is smaller), and so the lack of low eccentricity
companions at long periods could be due to this effect. In any case,
there are so few BD in this plot, that it is not clear whether \bds
low eccentricity is significant, so further comment on this will have
to wait until more data is available.

\begin{figure}[tbp]
\plotone{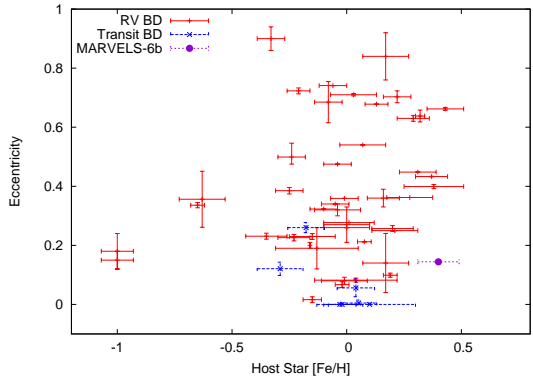}
\caption{The host metallicity and orbital eccentricity of \bd compared
  to the literature brown dwarfs from the catalog of \citet{ma13b} \bd
  has a relatively low eccentricity and its host star has a quite high
  [Fe/H].
\label{metalecc}
}
\end{figure}

This BD companion is also an outlier in metallicity space. In Figure
\ref{metalecc} \bd is again plotted with the BD from
\citet{ma13b}. \starnsp's metallicity is significantly super solar,
and \bd exists in an empty region of eccentricity-metallicity
space. As in the previous figure, the plot is merely suggestive of
interesting astrophysics, but there are not a sufficient number of BDs
to know whether \bd is a significant deviation. \bd is helping to fill
in the BD parameter space, and with more data its location may become
more astrophysically significant.

\section{Conclusion}
In this paper, we report the discovery of a BD companion to \star with
a minimum mass of 31.7 $\pm$ 2.0 $M_{Jup}$. This BD, designated \bdnsp,
has a moderately long period of $47.8929^{+0.0063}_{-0.0062}$ days
with a low eccentricty of $0.1442^{+0.0078}_{-0.0073}$.  We have
analyzed moderate resolution spectroscopy of the host star and have
determined the following properties: $T_{\rm eff} = 5598 \pm 63$,$\log
{g} = 4.44 \pm 0.17$, and [Fe/H] $= +0.40 \pm 0.09$. From these
measurements we find that \star has a mass and radius of $M_*= 1.11
\pm 0.11~M_\sun$ and $R_*= 1.06 \pm 0.23~R_\sun$. This result combined
with photometry, indicates the host star is a G dwarf at 219 $\pm$ 21
pc from the Sun with an age less than approximately 6 Gyr based on an
evolutionary track analysis.

Due to its moderately long period, \bd has a transit probability for a
uniform distribution of $\cos i$ of only 1.8\%. Although we have
roughly 13,000 photometric data points, we cannot conclusively rule
out a transit. In the Keck AO imaging no visual companions were
found. However, in the LI a previously known companion at
7.7\arcsec~from the host star was detected. This visual companion
appears in both the 2MASS and KIC catalogs, and was shown not be a
physical companion based upon photometry and astrometry, which is
unexpected given that many of the previous BD found by the MARVELS
survey did have tertiary companions.

Finally, we found that minimum mass of \bd exists at the minimum of
the mass functions of close (orbital period $< 5$ yr) stellar and
  planetary companions to stars, making this a rare object even
compared to other BDs. It also exists in an underdense region in both
period/eccentricity and metallicity/eccentricity space. This
ultimately furthers the goal of this series of papers, to help fill
out the low-mass companion phase space, which will ultimately help us
understand these intriguing objects.

The SuperWASP and Allegheny lightcurve data for \star, along with the
APO spectroscopic data will be made available through the
Vizier/CDS\footnote{http://vizier.u-strasbg.fr/viz-bin/VizieR} catalog
service. The SuperWASP and Allegheny lightcurve data will also be
available with the online edition of this article.



\acknowledgments
Funding for the MARVELS multi-object Doppler instrument was provided by the W.M. Keck Foundation and NSF with grant AST-0705139.
The MARVELS survey was partially funded by the SDSS-III consortium, NSF grant AST-0705139, NASA with grant NNX07AP14G and the University of Florida. 

Funding for SDSS-III has been provided by the Alfred P. Sloan Foundation, the Participating Institutions, the National Science Foundation, and the U.S. Department of Energy Office of Science. The SDSS-III web site is http://www.sdss3.org/.

SDSS-III is managed by the Astrophysical Research Consortium for the Participating Institutions of the SDSS-III Collaboration including the University of Arizona, the Brazilian Participation Group, Brookhaven National Laboratory, University of Cambridge, Carnegie Mellon University, University of Florida, the French Participation Group, the German Participation Group, Harvard University, the Instituto de Astrofisica de Canarias, the Michigan State/Notre Dame/JINA Participation Group, Johns Hopkins University, Lawrence Berkeley National Laboratory, Max Planck Institute for Astrophysics, Max Planck Institute for Extraterrestrial Physics, New Mexico State University, New York University, Ohio State University, Pennsylvania State University, University of Portsmouth, Princeton University, the Spanish Participation Group, University of Tokyo, University of Utah, Vanderbilt University, University of Virginia, University of Washington, and Yale University.

Based on observations made with the Italian Telescopio Nazionale Galileo (TNG) operated on the island of La Palma by the Fundación Galileo Galilei of the INAF (Istituto Nazionale di  Astrofisica) at the Spanish Observatoriodel Roque de los Muchachos of the Instituto de Astrofisica de Canarias.
The Center for Exoplanets and Habitable Worlds is supported by the
Pennsylvania State University, the Eberly College of Science, and the
Pennsylvania Space Grant Consortium.
Keivan Stassun, Leslie Hebb, and Joshua Pepper acknowledge funding support from the Vanderbilt Initiative in Data-Intensive Astrophysics (VIDA) from Vanderbilt University, and from NSF Career award AST-0349075.
BSG and JDE acknowledge support from NSF CAREER grant AST-1056524
EA thanks NSF for CAREER grant 0645416.
GFPM acknowledges financial support from CNPq grant n$^{\circ}$ 476909/2006-6 and FAPERJ grant n$^{\circ}$ APQ1/26/170.687/2004.
LG acknowledges financial support provided by the PAPDRJ CAPES/FAPERJ Fellowship


{\it Facilities:} \facility{Sloan ()}

\clearpage

\end{document}